\documentclass[submission,copyright,creativecommons]{class/eptcs}

\usepackage[T1]{fontenc}
\usepackage{amsmath}
\usepackage{amsthm}
\usepackage{amsfonts}
\usepackage{mathtools}
\usepackage{stmaryrd}
\usepackage[bb=boondox]{mathalfa}

\usepackage{graphicx}
\usepackage{multicol}
\usepackage{ifdraft}
\usepackage{ebproof}
\usepackage{algorithm}
\usepackage{algorithmicx}
\usepackage{xspace}
\usepackage[noend]{algpseudocode}
\usepackage{tikz, pgf}
\usepackage{hyperref}
\usepackage{url}
\usepackage[figbotcap]{subfigure}

\usepackage{ifthen}

\usepackage[short]{lib/latexutils/abbrev}
\usepackage{lib/latexutils/sidenote}
\usepackage{lib/latexutils/symbols}
\usepackage{lib/latexutils/theorems} 
\usepackage{zref-totpages}

\usepackage{appendix}

\usepackage[frozencache,cachedir=_minted]{local/minted}

\newcommand{\haskell}[1]{\mintinline{haskell}{#1}}

\setminted{
  linenos=true,
  autogobble
}

\ebproofset{
  right label template=\textsc{\inserttext}
}
\ebproofnewstyle{small}{
  separation = 1em, rule margin = .5ex,
  template = \small$\inserttext$,
  right label template=\scriptsize\textsc{\inserttext}
}

\usetikzlibrary{cd, arrows.meta, decorations.pathmorphing}
\tikzset{
  >=Kite,
  invisible/.style={opacity=0},
  visible on/.style={alt={#1{}{invisible}}},
  alt/.code args={<#1>#2#3}{%
      \alt<#1>{\pgfkeysalso{#2}}{\pgfkeysalso{#3}}%
  }
}
\tikzcdset{
  arrow style=tikz,
  diagrams={>=Kite},
  ampersand replacement=\&,
  labels=description
}

\algrenewcommand\algorithmicindent{1.0em}%
\algrenewcommand\algorithmicfunction{\textbf{fn}}
\algnewcommand\algorithmicyield{\textbf{yield}\ }
\algnewcommand\algorithmiccontinue{\textbf{continue}}
\algnewcommand\algorithmicbreak{\textbf{break}}

\DeclareMathOperator*{\powerset}{\ensuremath{\mathcal{P}}}
\newcommand{\tstr}{\ensuremath{\mathtt{str}}}

\newcommand\omstarr{\ensuremath{\omega^{\star}_r}}
\newcommand\omr{\ensuremath{\omega_r}}

\ifthenelse{\boolean{true}}{%
\newcommand\freetrR[1]{\ensuremath{\xmapsto[r\star]{#1}}}
\renewcommand\freetr[1]{\ensuremath{\xmapsto[a\star]{#1}}}
\newcommand\trR[1]{\ensuremath{\xmapsto[r]{#1}}}
\newcommand\trA[1]{\ensuremath{\xmapsto[a]{#1}}}
}{%
\newcommand\freetrR[1]{\ensuremath{\xmapsto[\omstarr]{#1}}}
\renewcommand\freetr[1]{\ensuremath{\xmapsto[\omstara]{#1}}}
\newcommand\trR[1]{\ensuremath{\xmapsto[\omega_r]{#1}}}
}

\renewcommand\fembed{\ensuremath{\mathit{embed}}}

\newcommand\abstractr{\ensuremath{\mathit{abstract}_r}}

\newboolean{preprint}

\theoremstyle{plain}
\newtheorem*{theorem*}{Theorem}

\setboolean{preprint}{false}
\setboolean{submission}{true}

\providecommand{\event}{ICLP 2025}

\title{A Refined Operational Semantics for FreeCHR}
\author{Sascha Rechenberger and Thom Frühwirth
\institute{%
Institute for Software Engineering and Programming Languages\\
Ulm University\\
Albert-Einstein-Allee 11, 89069 Ulm, Germany}
\email{\{sascha.rechenberger,thom.fruehwirth\}@uni-ulm.de}
}

\newcommand{\titlerunning}{A Refined Operational Semantics for FreeCHR}
\newcommand{\authorrunning}{S. Rechenberger, T. Frühwirth}

\hypersetup{
    bookmarksnumbered,
    pdftitle    = {\titlerunning},
    pdfauthor   = {\authorrunning},
    pdfsubject  = {EPTCS},               
    pdfkeywords = {%
        embedded domain-specific languages,%
        rule based programming languages,%
        constraint handling rules,%
        operational semantics,%
        initial algebra semantics%
    } 
}

\begin{document}
  \maketitle
  \begin{abstract}
  \emph{Constraint Handling Rules} (CHR) is a rule-based programming language which is typically embedded into a general-purpose language.
  There exists a plethora of implementations for numerous host languages.
  However, the existing implementations often re-invent the way to embed CHR, which impedes maintenance and weakens assertions of correctness.
  To formalize and thereby standardize the embedding into arbitrary host languages, we introduced the framework \emph{FreeCHR} and proved it to be a valid representation of classical ground CHR.
  Until now, this framework only includes a translation of the \emph{very abstract} operational semantics which, due to its abstract nature, is not a sufficient base for practical implementations.
  In this paper we present a translation of the \emph{refined} operational semantics for FreeCHR and prove it to be both a valid concretization of the \emph{very abstract} semantics of FreeCHR, and an equivalent representation of the \emph{refined} semantics of CHR.
  This will establish implementations of FreeCHR as equivalent in behavior and expressiveness to existing implementations of CHR.
  \ifthenelse{\boolean{preprint}}{This is an extended preprint of a technical report submitted to \emph{the 41st International Conference on Logic Programming}.}{}
\end{abstract}

  \ifdraft{\tableofcontents}{}

  \section{Introduction}\label{sec::freechr:introduction}
\emph{Constraint Handling Rules} (CHR) is a rule-based programming language which is typically embedded into a general-purpose language.
Having a CHR implementation available enables software developers to solve problems in a declarative and elegant manner.
Aside from the obvious task of implementing constraint solvers \cite{fruehwirth2006complete,dekoninck2006inclp}, it has been used, \EG to solve scheduling problems \cite{abdennadher2000university} and implement concurrent and multi-agent systems \cite{thielscher2002reasoning,thielscher2005flux,lam2006agent,lam2007concurrent}.
In general, CHR is ideally suited for any problem that involves the transformation of collections of data.
Programs consist of rewriting rules which hide the process of finding matching values to which rules can be applied.
As a result, developers are able to implement algorithms and applications in a purely declarative way, without the otherwise necessary boilerplate code.

The literature on CHR as a formalism consists of a rich body of theoretical work, including a rigorous formalization of its declarative and operational semantics \cite{fruehwirth2009constraint,sneyers2010time,fruehwirth2015constraint}, relations to other rule-based formalisms \cite{fruehwirth2025principles} and results on properties like confluence \cite{abdennadher1996confluence,christiansen2015confluence,gall2017decidable}.
Implementations of CHR exist for a number of languages, such as Prolog \cite{schrijvers2004leuven}, C \cite{wuille2007cchr}, C++ \cite{barichard2024chr}, Haskell \cite{chin2008typesafe,lam2007concurrent}, JavaScript \cite{nogatz2018chr} and Java \cite{abdennadher2002jack,vanweert2005leuven,ivanovic2013implementing,wibiral2022javachr}.

We intend to formalize and thereby standardize the embedding of CHR into arbitrary host languages as an internal DSL.
In our view, this will have three major advantages.
First, by formalizing CHR via \emph{initial algebra semantics}, we provide a tighter connection between the formal definition and the implementations.
Initial algebra semantics are a common concept in functional programming and are used to inductively define and implement languages and their semantics \cite{hudak1998modular,johann2007initial}.
This provides a common formal foundation for considerations concerning theory and implementations.

Secondly, by embedding CHR as an internal language, distribution and use are greatly increased.
Unlike embeddings which rely on an external source-to-source compiler, internal embeddings can be easily distributed and imported via the host languages package and module system.
They also do not add another link to the build chain which is, even with modern build tools, an additional nuisance. 

Finally, by providing formal, host language agnostic and verifiable definitions of a CHR implementation, we also provide a generalized documentation for implementations.
We hope that this will ease contribution to (in future) existing implementations and help to keep them maintained.

We hence introduced the framework \emph{FreeCHR} \cite{rechenberger2025freechr}.
It provides both, a guideline and high-level architecture to implement and maintain CHR implementations across host languages, and a strong connection between the practical and formal aspects of CHR.
By using initial algebra semantics, we also get a formalization for type-safe programs as well as composition of programs for free.

However, the framework currently only formalizes the \emph{very abstract} operational semantics of CHR which are too abstract to serve as basis for standardized efficient implementations.
The \emph{refined} operational semantics $\omr$, introduced by Duck et al. \cite{duck2004refined}, formalizes the behavior of efficient existing CHR systems.
It fixes the order in which rules are tried to be applied as well as the order in which the head of a rule is traversed in order to find suitable values for rule application.
Hereby, the programmer can use the control flow of the CHR system to write more efficient programs.
The refined semantics also prevents reapplication of rules with the same applicable values.
This fixes the issue of trivial non-termination of certain types of programs.

In this paper we hence will introduce the \emph{refined} semantics to FreeCHR.
We will provide a modified definition $\omstarr$ and prove their soundness and completeness \WRT $\omr$.
We will also prove soundness \WRT the existing \emph{very abstract} operational semantics to guarantee that all definitions within our framework are consistent.
The operational semantics presented herein are already implemented in our ongoing work \cite{rechenberger2025instance,rechenberger2025optimized}.

The rest of the paper is structured as follows:
\autoref{sec:functors} introduces necessary preliminary notations and definitions, \autoref{sec:non-herbrand-chr} the definition of ground \emph{Constraint Handling Rules} over non-Herbrand domains and the \emph{refined} semantics for CHR, and \autoref{sec:freechr} the definition of \emph{FreeCHR} and its \emph{very abstract} semantics.
The remaining sections contain our new contributions.
\autoref{sec:freechr:refined} introduces the \emph{refined} semantics of FreeCHR, \autoref{sec:freechr:refined:va} proves their soundness \WRT the \emph{very abstract} semantics of FreeCHR and \autoref{sec:freechr:refined:refined} proves their soundness and completeness \WRT the \emph{refined} semantics of CHR.
Finally \autoref{sec:conclusion} concludes the paper.
\ifthenelse{\boolean{submission}}{An extended preprint of this paper, including full proofs, is available at \url{arXiv.com} \cite{rechenberger2025refined}.}{}

  \section{Preliminaries}\label{sec:functors}
We first want to define the basic concepts and notations we use throughout the paper.

\subsection{Basic notations}
The \emph{disjoint union} $ A \sqcup B = \left\{l_A(a) \mid a \in A\right\} \cup \left\{l_B(b) \mid b \in B\right\}$ of two sets $A$ and $B$ is the union of both sets, with additional labels $l_A$ and $l_B$ added to the elements, to keep track of the origin set of each element.
We will also use the labels $l_A$ and $l_B$ as \emph{injection} functions $l_A : A \rightarrow A \sqcup B$ and $l_B : B \rightarrow A \sqcup B$ which construct elements of $A \sqcup B$ from elements of $A$ or $B$, respectively.
For two functions $f : A \rightarrow C$ and $g : B \rightarrow C$, the function $\left[f,g\right] : A \sqcup B \rightarrow C$ is defined as
\begin{align*}
  \left[f,g\right](l(x)) &=\left\{\begin{matrix*}[l]
    f(x), & & \mbox{if}\ l = l_A \\
    g(x), & & \mbox{if}\ l = l_B \\
  \end{matrix*}\right.
\end{align*}
It can be understood as a formal analogue to a \haskell{case ... of} expression.
Furthermore, we define two functions $f \sqcup g : A \sqcup B \rightarrow A' \sqcup B'$ and $f \times g : A \times B \rightarrow A' \times B'$ as
\begin{multicols}{2}
  \noindent
  \begin{align*}
    (f \sqcup g)(l(x)) &= \left\{
      \begin{matrix*}[l]
        l_{A'}(f(x)), & &\mbox{if}\ l = l_A \\
        l_{B'}(g(x)), & &\mbox{if}\ l = l_B
      \end{matrix*}
    \right.
  \end{align*}
  \begin{align*}
    (f \times g)(x,y) &= (f(x), g(y))
  \end{align*}
\end{multicols}
\noindent
which lift two functions $f : A \rightarrow A'$ and $g : B \rightarrow B'$ to the disjoint union or the Cartesian product, respectively.

\subsection{Functors and $F$-algebras}
A functor $F$ maps all sets $A$ to sets $F A$ and all functions $f : A \rightarrow B$ to functions $F f : F A \rightarrow F B$, 
such that $F \id_A = \id_{F A}$ and $F (g \circ f) = F g \circ F f$. $\id_X(x) = x$ is the identity function on a set $X$.
A signature $\Sigma = \left\{\sigma_1/a_1,...,\sigma_n/a_n\right\}$, where $\sigma_i$ are operators and $a_i$ their arity, generates a functor $F_{\Sigma}$ with
\begin{align*}
  F_{\Sigma} X = \bigsqcup_{\sigma/a\in\Sigma}\ X^a && F_{\Sigma} f = \bigsqcup_{\sigma/a\in\Sigma}\ f^a
\end{align*}
$X^{n}$ and $f^{n}$ are defined as
\begin{align*}
  X^{n} = \underbrace{X \times ... \times X}_{\text{$n$ times}} && f^{n} = \underbrace{f \times ... \times f}_{\text{$n$ times}}
\end{align*}
and $X^0 = \mathbb{1}$ and $f^0 = \id_{\mathbb{1}}$, where $\mathbb{1}$ is a singleton set.
Such a functor $F_{\Sigma}$ models \emph{flat} (\IE, not nested) terms over the signature $\Sigma$.

Since an endofunctor $F$ defines the syntax of terms, an evaluation function $\alpha : F A \rightarrow A$ defines the \emph{semantics} of terms.
We call such a function $\alpha$, together with its \emph{carrier} $A$, an $F$-algebra $(A, \alpha)$.
If there are two $F$-algebras $\left(A, \alpha\right)$ and $\left(B, \beta\right)$ and a function $h : A \rightarrow B$, we call $h$ an \emph{$F$-algebra homomorphism}, \IFF $h \circ \alpha = \beta \circ F h$, i.e., $h$ preserves the structure of $\left(A, \alpha\right)$ in $\left(B, \beta\right)$ when mapping $A$ to $B$.
In this case, we also write $h : \left(A,\alpha\right) \rightarrow \left(B, \beta\right)$.

A special $F$-algebra is the \emph{free $F$-algebra} $F^{\star} = (\mu F, \cons_F)$, for which there is a homomorphism\linebreak$\kata{\alpha} : F^{\star} \rightarrow \left(A, \alpha\right)$ for any other algebra $\left(A, \alpha\right)$.
We call those homomorphisms $\kata{\alpha}$ \emph{$F$-algebra catamorphisms}.
The functions $\kata{\alpha}$ encapsulate structured recursion on values in $\mu F$ with the semantics defined by the function $\alpha$ which is itself only defined on flat terms.
The carrier of $F^{\star}$, with $\mu F = F \mu F$, is the set of inductively defined values in the shape defined by $F$.
The function $\cons_F : F \mu F \rightarrow \mu F$ inductively constructs the values in $\mu F$.

\subsection{Labelled transition systems}
A \emph{labelled transition system} (LTS) $\omega = \langle \Sigma, L, (\mapsto) \rangle$
consists of a set $\Sigma$ called the \emph{domain},
a set $L$ called \emph{actions}
and a ternary \emph{transition relation} $\left(\mapsto\right) \subseteq \Sigma \times L \times \Sigma$.
The idea is that if $s \xmapsto{\ l\ } s' \in \left(\mapsto\right)$, we transition from state $s$ to $s'$ by the action $l$.

For two LTS $\omega_1 = \langle \Sigma_1, L_1, (\mapsto) \rangle$ and $\omega_2 = \langle \Sigma_2, L_2, (\hookrightarrow) \rangle$
and two functions $f : \Sigma_1 \longrightarrow \Sigma_2$ and\linebreak$g : L_1 \longrightarrow L_2$ we say that $\omega_1$ is $(f, g)$-sound \WRT $\omega_2$, \IFF
\begin{align*}
  s \xmapsto{\ l\ } s' \in (\mapsto) \Longrightarrow f(s) \xhookrightarrow{g(l)} f(s') \in (\hookrightarrow)\tag{$(f, g)$-soundness}
\end{align*}
and
$(f,g)$-complete \WRT $\omega_2$, \IFF
\begin{align*}
  s \xmapsto{\ l\ } s' \in (\mapsto) \Longleftarrow f(s) \xhookrightarrow{g(l)} f(s') \in (\hookrightarrow)\tag{$(f, g)$-completeness}
\end{align*}
By $(\mapsto)^{+}$ we denote the \emph{transitive} and by $(\mapsto)^{*}$ the \emph{reflexive-transitive} closure of $(\mapsto)$.
Recall that $(\mapsto)^{+} \subset (\mapsto)^{*}$, for every $R \subseteq (\mapsto)^{+}$, $R^{+} \subseteq (\mapsto)^{+}$ and
for every $Q \subseteq (\mapsto)^{*}$, $Q^{*} \subseteq (\mapsto)^{*}$.

  \section{Ground CHR over non-Herbrand domains}\label{sec:non-herbrand-chr}
In this section, we will reiterate definitions concerning the generalization of CHR to non-Herbrand domains and introduce its \emph{refined} operational semantics.

A \emph{data type} determines the syntax and semantics of terms via a functor $\Lambda_T$ and a $\Lambda_T$-algebra $\tau_T$.
The fixed point $\mu\Lambda_T$ contains terms which are inductively defined via $\Lambda_T$ and the $\Lambda_T$-catamorphism $\kata{\tau_T}$ evaluates those terms to values of $T$.

\begin{definition}[Data types]\label{def::lan:data-type}
  A \emph{data type} is a triple $\langle T, \Lambda_{T}, \tau_T\rangle$, where $T$ is a set, $\Lambda_T$ a functor and $\left(T, \tau_T\right)$ a $\Lambda_T$-algebra.
\end{definition}

We write $t \equiv_{T} t'$ for $t \in \mu\Lambda_T$ and $t' \in T$, \IFF $\kata{\tau_T}(t) = t'$. 

\begin{example}[Boolean data type]
  The signature
  \begin{align*}
    \Sigma_{\tbool} =& \left\{(n \leq m)/0 \mid n,m\in\mathbb{N}_0\right\}
      \cup \left\{(n < m)/0 \mid n,m\in\mathbb{N}_0\right\}
      \cup \left\{\band/2, \btrue/0, \bfalse/0\right\}
  \end{align*}
  defines Boolean terms\footnote{We will generally overload symbols like $\bfalse$, $\btrue$, $\band$, $\bnot$, ..., if their meaning is clear from the context.}.
  $\Sigma_{\tbool}$ generates the functor
  \begin{align*}
    \Lambda_{\tbool} X =&\ \mathbb{N}_0 \times \mathbb{N}_0
     \sqcup\ \mathbb{N}_0 \times \mathbb{N}_0
     \sqcup\ X \times X \sqcup \mathbb{1} \sqcup \mathbb{1}
  \end{align*}
  the fixed point, $\mu\Lambda_{\tbool}$, of which is the set of valid nested Boolean terms like $\left(0 < 4 \band 4 \leq 6\right)$.
  Let $\langle \tbool, \Lambda_{\tbool}, \tau_{\tbool} \rangle$, with $\tbool = \left\{\btrue, \bfalse\right\}$, be a data type. 
  If we assume $\tau_{\tbool}$ to implement the usual semantics for Boolean terms and comparisons, $\left(0 < 4 \band 4 \leq 6\right)$ will evaluate as 
  \begin{align*}
    \kata{\tau_{\tbool}}(0 < 4 \band 4 \leq 6) = \kata{\tau_{\tbool}}(0 < 4) \band \kata{\tau_{\tbool}}(4 \leq 6) = \btrue \band \btrue = \btrue
  \end{align*}
  We can therefor write $0 < 4 \band 4 \leq 6 \equiv_{\tbool} \btrue$.
\end{example}

For a set $T$, both $\Lambda_T$ and $\tau_T$ are determined by the host language which is captured by the next definition.
\begin{definition}[Host environment]\label{def::lan:host-environment}
  A mapping $\mathcal{L} T = \langle T, \Lambda_T, \tau_T \rangle$, where $\langle T, \Lambda_T, \tau_T \rangle$ is a data type, is called a \emph{host environment}.
\end{definition}

We assume that a host environment is provided by the host language (and the program, the CHR program is part of) and assigns a data type to a set $T$, effectively determining syntax and semantics of terms that evaluate to values of $T$.

\begin{definition}[Non-Herbrand ground CHR programs]\label{def::chr:syn:program}
  CHR programs are sequences of multiset-rewriting rules of the generalized form 
  \begin{align*}
    &N\ @\ K\ \setminus\ R\ \Longleftrightarrow\ \left[G\ |\right]\ B
  \end{align*}
  $N$ is the identifier of the rule.
  It has to be unique in the executed program.
  For a set $C$, called the \emph{domain} of the Program, for which there is a data type $\mathcal{L} C = \langle C, \Lambda_C, \tau_C \rangle$,
  $K, R \in \flist C$ are called the \emph{kept} and \emph{removed head}, respectively.
  Either can be omitted, but not both at the same time.
  If $K$ is empty, we call the rule a \emph{simplification} rule\ifthenelse{\boolean{preprint}}{.
  If $R$ is empty, we call it a \emph{propagation} and write them with $(\Longrightarrow)$ instead of $(\Longleftrightarrow)$}{, and if $R$ is empty, a \emph{propagation} rule}.
  If both $K$ and $R$ are non-empty, the rule is called a \emph{simpagation} rule.
  The functor $\flist X = \bigcup_{i\in\mathbb{N}} X^i$ maps a set $X$ to the set of finite sequences (or lists) over X, with $X^0 =[]$ being the empty sequence\footnote{We use Haskell-like syntax to denote lists or sequences. $[]$ is the empty list and $\mathit{x}:\mathit{xs}$ constructs a list with head element $\mathit{x}$ and tail $\mathit{xs}$. We will also use the notations $\left(a : b : c : []\right)$ and $\left[a, b, c\right]$ interchangeably as we consider it useful. The operator $(\diamond) : \flist C \times \flist C \rightarrow \flist C$ denotes list concatenation.}.
  The optional $G \in \mu\Lambda_{\tbool}$ is called the \emph{guard}. If $G$ is omitted, we assume $G \equiv_{\tbool} \btrue$.
  $B \in \flist\mu\Lambda_C$ is called the \emph{body}.
\end{definition}

The members of the kept and removed head are matched against values of the domain $C$.
The guard $G$ is a term that can be evaluated to a Boolean value in $\tbool$.
The body $B$ is a list of terms which can be evaluated to values of $C$.
This includes any call of (pure) functions or operators which evaluates to Booleans, or values of $C$, respectively. 

\autoref{def::chr:syn:program} corresponds to the \emph{positive range-restricted ground} segment of CHR which is commonly used as the target for embeddings of other (rule-based) formalisms \cite[Chapter 7]{fruehwirth2025principles}.
$\prgc$ denotes the set of all such programs over a domain $C$.

\begin{example}[Euclidean algorithm]\label{ex:chr:gcd}
   The program $\textsc{gcd}=\left[\mathit{zero} @ ..., \mathit{subtract} @ ...\right]$
  \begin{align*}
    \mathit{zero}\ &@\ 0\ \Longleftrightarrow\ \emptyset\\
    \mathit{subtract}\ &@\ N\ \setminus\ M\ \Longleftrightarrow\ 0 < N \band 0 < M \band N \leq M\ |\ M-N
  \end{align*}
  computes the greatest common divisor of a collection of natural numbers.
  The \emph{zero} rule removes any occurrences of $0$ from the collection.
  The \emph{subtract} rule replaces for any pair of numbers $N$ and $M$ greater $0$ and $N \leq M$, $M$ by $M-N$.
\end{example}

\begin{definition}[$C$-groundings \& instances of rules]\label{def::chr:syn:pro:instance}
  For a \emph{positive range-restricted ground} rule
  \begin{align*}
    r = R\ @\ k_1,...,k_n\ \setminus\ r_{1},...,r_{m}\ \Leftrightarrow\ G\ |\ B
  \end{align*}
  with universally quantified variables $v_1,...,v_l$,
  and a data type $\mathcal{L} C = \langle C, \Lambda_{C}, \tau_{C} \rangle$, 
  we call the set
  \begin{align*}
    \Gamma_{C}(r) =
      \{\ &(R\ @\ k_1\sigma,...,k_n\sigma\ \setminus\ r_{1}\sigma,...,r_{m}\sigma\ \Leftrightarrow\ G\sigma\ |\ \flist\kata{\tau_{C}}(B\sigma)) \\
      \mid\ &\mbox{the substitution}\ \sigma\ \mbox{instantiates all variables $v_1$, ..., $v_l$},\\
      \ &k_1\sigma,...,k_n\sigma, r_1\sigma,...,r_m\sigma \in C,\\
      \ &G \sigma \in \mu\Lambda_{\tbool},\\
      \ &B \sigma \in \flist\mu\Lambda_C\ \}
  \end{align*}
  the \emph{$C$-grounding} of $r$.
  Analogously, for a program $\mathcal{R}$, $\Gamma_C(\mathcal{R}) = \bigcup_{r \in \mathcal{R}} \Gamma_C(r)$ is the $C$-grounding of $\mathcal{R}$.
  An element $r' \in \Gamma_{C}(r)$ (or $\Gamma_C(\mathcal{R})$ respectively) is called a \emph{$C$-instance} of a rule $r \in \mathcal{R}$.
\end{definition}

A C-instance (or grounding) is obtained, by instantiating any variables and evaluating the then ground terms in the body of the rule, using the $\Lambda_C$-catamorphism $\kata{\tau_C}$.
\begin{example}\label{ex::freechr:nhc:body-instance}
Given a body $\left[M - N\right]$ and a substitution $\sigma = \left\{N \mapsto 4, M \mapsto 6\right\}$, the body is instantiated like
\begin{align*}
  \flist\kata{\tau_C}(\left[M-N\right]\sigma) = \flist\kata{\tau_C}(\left[6-4\right]) = \left[\kata{\tau_C}(6-4)\right] = \left\{2\right\}
\end{align*}
\end{example}
With \autoref{ex::freechr:nhc:body-instance}, we can also easily see that if we use a data type $\mathcal{L}\mu\Lambda_C = \langle \mu\Lambda_C, \Lambda_C, \cons_{\Lambda_C} \rangle$ we get the Herbrand interpretation of terms over $C$.
Hence, \FI an expression $\left(3+4\right) \in \mu\Lambda_{\mathbb{N}_0}$ is evaluated to itself, as it is the case in Prolog.

\begin{example}[$C$-instances]\label{ex:chr:gcd:instances}
  If we instantiate the rule
  \begin{align*}
    subtract\ &@\ N\ \setminus\ M\ \Leftrightarrow\ 0 < N \band 0 < M \band N \leq M\ |\ M-N
  \end{align*} 
  with $\sigma_1 = \left\{N \mapsto 4, M \mapsto 6\right\}$ and $\sigma_2 = \left\{N \mapsto 0,M \mapsto 6\right\}$, respectively,
  we get the $\mathbb{N}_0$-instances
  \begin{align*}
    (subtract) \sigma_1 = subtract\ &@\ 4\ \setminus\ 6\ \Leftrightarrow\ 0 < 4 \band 0 < 6 \band 4 \leq 6\ |\ 2\\
    (subtract) \sigma_2 = subtract\ &@\ 0\ \setminus\ 6\ \Leftrightarrow\ 0 < 0 \band 0 < 6 \band 0 \leq 6\ |\ 6
  \end{align*}
  Both instances are elements of the $\mathbb{N}_0$-grounding $\Gamma_{\mathbb{N}_0}\left(\textsc{gcd}\right)$ of the program in \autoref{ex:chr:gcd}.
\end{example}

Classically, the guard $G$ contains constraints which are defined \WRT a constraint theory $\mathcal{CT}$.
We typically write $\mathcal{CT} \models G$ to denote that the guard is satisfiable \WRT $\mathcal{CT}$.
In our case, $\mathcal{CT}$ is essentially defined by $\tau_{\tbool}$, as it determines the semantics of Boolean terms.
We thus write $\tau_{\tbool} \models G$ \IFF  $G \equiv_{\tbool} \btrue$ and $\tau_{\tbool} \models \bnot G$ \IFF $G \equiv_{\tbool} \bfalse$.
Note that we always need a data type $\mathcal{L} \tbool$.
\begin{example}[$\oma$-transitions]\label{ex::chr:sem:op:very-abstract}
  Given the instances from \autoref{ex:chr:gcd:instances}, we write 
  \begin{align*}
    \tau_{\tbool} \models 0 < 4 \band 0 < 6 \band 4 \leq 6
  &&\text{and}&&
    \tau_{\tbool} \not\models  0 < 0 \band 0 < 6 \band 0 \leq 6
  \end{align*}
\end{example}

The very abstract operational semantics $\oma$ of CHR operates on plain multisets of values.
It describes that a rule $r$ can be applied to a state $\left\{k_1, ..., k_n, r_1, ..., r_m\right\} \uplus \Delta s$ if there is a $C$-instance
\begin{align*}
  \left(k_1, ..., k_n\ \setminus\ r_1, ..., r_m\ \Longleftrightarrow\ \left.G \mid b_1, ..., b_p\right.\right) \in \Gamma_C(r)
\end{align*}
with $G \equiv_{\tbool} \btrue$.
If the rule is applied, the elements $\left\{r_1, ..., r_m\right\}$ are replaced by $\left\{b_1, ..., b_p\right\}$.
How the \emph{matching} $\left\{k_1, ..., k_n, r_1, ..., r_m\right\}$ is found and which one of the applicable rules of the executed program is applied is non-deterministic \cite{fruehwirth2009constraint}.
The following example shows the execution of the Euclidean algorithm as a final example of the operational semantics of CHR.

\begin{example}[Euclidean algorithm (cont.)]
  The rules of \textsc{gcd} are applied until exhaustion, leaving only the greatest common divisor of all numbers of the input.
  For an input $\left\{4,6\right\}$, the program will perform a sequence
  \begin{align*}
    \left\{4,6\right\}
    \trA{\left\{\mathit{subtract}\right\}} \left\{4,2\right\}
    \trA{\left\{\mathit{subtract}\right\}} \left\{2,2\right\}
    \trA{\left\{\mathit{subtract}\right\}} \left\{2,0\right\}
    \trA{\left\{\mathit{zero}\right\}} \left\{2\right\}
  \end{align*}
  of transformations.
\end{example}

The very abstract semantics also does not account for propagation rules which remain applicable to a state, since none of the applicable values are removed.
This possibly leads to non-termination if a program contains propagation rules.

\begin{example}[Transitive hull]\label{ex:chr:trans}
  Given the program
  \begin{align*}
    \mathit{trans}\ @\ \left(X, Y\right)^{\#2},\ \left(Y, Z\right)^{\#1}\ \Longrightarrow\ X \neq Z\ |\ \left(X, Z\right)
  \end{align*}
  which adds the transitive edge $(X, Z)$ of two edges $(X,Y)$ and $(Y, Z)$, with $X \neq Z$.

  Since reapplication of a rule to the same values is allowed under the very abstract semantics the infinite transition sequence
  \begin{align*}
    &\left\{\left(a, b\right), \left(b, c\right)\right\}\\
    \trA{\left\{\mathit{trans}\right\}} &\left\{\left(a, b\right), \left(b, c\right), \left(a, c\right)\right\}\\
    \trA{\left\{\mathit{trans}\right\}} &\left\{\left(a, b\right), \left(b, c\right), \left(a, c\right), \left(a, c\right)\right\}\\
    \trA{\left\{\mathit{trans}\right\}} &\left\{\left(a, b\right), \left(b, c\right), \left(a, c\right), \left(a, c\right), \left(a, c\right)\right\}\\
    \trA{\left\{\mathit{trans}\right\}} &\ ...
  \end{align*}
  is perfectly valid but would practically lead to non-termination.
\end{example}

The refined semantics solves these issues by resolving some sources of non-determinism and guaranteeing that rules are not applied more than once with the same matching.
However, it requires additional information in the states to do so.

\begin{definition}[States]\label{def:states}
  The functor
  \begin{align*}
      \Omega_r C = \flist (C \setsum (\tnat \setprod C))
           \;\setprod\; \powerset (\tnat \setprod C)
           \;\setprod\; \powerset (\tstr \setprod \flist \tnat)
           \;\setprod\; \tnat
  \end{align*}
  models the set of states over values in $C$.
  For an element $\langle Q, S, H, I \rangle \in \Omega_r C$ we call
  $Q$ the \emph{query}, $S$ the \emph{store}, $H$ the \emph{propagation history} and $I$ the \emph{index}.
  $\powerset\left(X\right)$ is the powerset of a set $X$, $\tnat$ are the natural numbers and $\tstr$ are strings. 
\end{definition}
The \emph{query} can be understood as an execution stack.
If a value in the body of a rule is added, it is handled as if it was a procedure call, where the body of the procedure is defined by the rules of the program.
The \emph{store} is a set of identifier-value pairs.
It is classically viewed as a conjunction of currently known facts.
The unique identifier also serves the purpose of simulating the multiset semantics required by CHR and FreeCHR.
The \emph{propagation history} is used to prevent trivial non-termination.
Finally, the \emph{index} is a natural number that is used to generate unique identifiers.

In programs for the refined semantics, the head patterns are viewed as decorated with indices incrementing from \emph{right to left} and \emph{top to bottom} (in textual order) throughout the program.
We call them \emph{pattern indices}.
$\prgc^{\#}$ denotes the set of programs in $\prgc$ with enumerated patterns.
The function $\mathit{indices}$ maps a program in $p \in \prgc^{\#}$ to the set of pattern indices occurring in $p$.

\begin{example}[Greatest common divisor, enumerated]\label{ex:chr:gcd:decorated}
  The program $\textsc{gcd}^{\#}$ below computes the greatest common divisor.
  It is the enumerated variant of the program $\textsc{gcd}$ shown in \autoref{ex:chr:gcd}.
  The patterns of the rule are decorated with pattern indices.
  $\mathit{indices}(\textsc{gcd}^{\#})$ returns the set $\left\{1,2,3\right\}$.
    \begin{align*}
      \mathit{zero}\ &@\ 0^{\#1}\ \Longleftrightarrow\ \emptyset\\
      \mathit{subtract}\ &@\ N^{\#3}\ \setminus\ M^{\#2}\ \Longleftrightarrow\ 0 < N \band 0 < M \band N \leq M\ |\ M-N
    \end{align*}
\end{example}

We now define the \emph{refined} semantics of CHR.
The original definition describes six kinds of state transitions \cite{duck2004refined,fruehwirth2009constraint}.
Since we want to operate on ground values only, we can ignore two of them which are only concerned with non-ground values.

\begin{definition}[\emph{Refined} operational semantics for non-Herbrand ground CHR]
The \emph{refined} operational semantics for non-Herbrand ground CHR are defined as an LTS 
\begin{align*}
  \omr = \langle\;\Omega_r C\; ,\; \prgc^{\#}\; ,\; (\trR{})^{+}\; \rangle
\end{align*}
where the transition relation $(\trR{}) \subseteq \Omega_r C \times \prgc^{\#} \times \Omega_r C$ is defined by the rules below.

\paragraph*{\textbf{Activate}}
The transition
\begin{align*}
    \langle c:Q, S, H, I \rangle
    \trR{\ p\ } \langle (I, c)^{\#1} : Q, \left\{\left(I, c\right)\right\} \uplus S, H, I+1 \rangle \tag{\textsc{activate}}
\end{align*}
activates a value $c$ by introducing it to the store with a unique identifier $I$.
On the query, the value is also decorated with the pattern index $\#1$.
This indicates that it will be tried to match it to the rightmost pattern of the first rule of the program $p$.
The activation of a value can be understood as a procedure call where the procedure is defined by the applicable rules.
We use the operator $(\uplus)$ to emphasize that the operands are disjoints sets, \IE if $A \uplus B = C$, then $A \cup B = C$ and $A \cap B = \emptyset$.

\paragraph*{\textbf{Apply}}
Given a $C$-instance
\begin{align*}
  (N\ &@\ c_1^{\#l_1}, ..., c_{n}^{\#l_n}\ \setminus\ c_{n+1}^{\#l_{n+1}}, ..., c_{n+m}^{l_{n+m}}\ \Longleftrightarrow\ G\ |\ B ) \in \Gamma_C(p) 
\end{align*}
such that for $j \in \left\{1, ..., n+m\right\}$, $(i_j, c_j) \in K \uplus R$, $G \equiv_{\tbool} \btrue$ and $\left\{(N, i_1, ..., i_{n+m})\right\} \notin H$, we can perform the transition
\begin{align*}
  \langle (i_j, c_j)^{\#l_j} : Q, K \uplus R \uplus S, H, I\rangle
  \trR{\ p\ }
  \langle B \diamond ((i_j, c_j)^{\#l_j} : Q), K \uplus S, \left\{(N, i_1, ..., i_{n+m})\right\} \cup H, I \rangle \tag{\textsc{apply}}
\end{align*}
with $K=\left\{\left(i_1, c_1\right), ..., \left(i_n, c_n\right)\right\}$ and $R=\left\{\left(i_{n+1}, c_{n+1}\right), ..., \left(i_{n+m}, c_{n+m}\right)\right\}$.

We need to check if the configuration $(N, i_1, ..., i_{n+m})$ already fired to prevent reapplication.
If not, we record the configuration, remove $R$ from the store and query the values of the body, by concatenating the sequence $B$ before the query.

\paragraph*{\textbf{Drop}}
The transition
\begin{align*}
  \langle (i, c)^{\#j} : Q, S, H, I \rangle \trR{\ p\ } \langle Q, S, H, I \rangle \tag{\textsc{drop}}
\end{align*}
is used if $j$ exeeds the pattern indices of the program $p$, \IE if $j \notin \mathit{indices}(p)$. This indicates that there are no more applicable rules for the currently active value.
This also happens, if $(i,c)$ was removed by the \textsc{apply} transition at some point.

\paragraph*{\textbf{Default}}
If there is a rule $r \in p$, such that $j \in \mathit{indices}(r)$, but no $C$-instance of $r$, such that the \emph{Apply} transition can be used,
we use the transition
\begin{align*}
  \langle (i, c)^{\#j} : Q, S, H, I \rangle \trR{\ p\ } \langle (i, c)^{\#j+1} : Q, S, H, I \rangle \tag{\textsc{default}}
\end{align*}
This transition continues the traversal with the currently active value through the program.
\end{definition}

\ifthenelse{\boolean{preprint}}{%
We now want to demonstrate the \emph{refined} semantics on two examples.
The first example computes the greatest common divisor of 6 and 9 using the program from \autoref{ex:chr:gcd:decorated}.
The second computes the transitive hull of a simple graph $\langle \left\{a, b, c\right\}, \left\{(a, b), (b, c)\right\} \rangle$.

\begin{figure}[t]
  {\footnotesize\begin{align*}
                       &\ \langle [6, 9], \emptyset, \emptyset, 1 \rangle\\
    \trR{\textsc{gcd}} &\ \langle [(1, 6)^{\#1}, 9], \left\{(1, 6)\right\}, \emptyset, 2 \rangle\tag{\textsc{activate}}\\
    \trR{\textsc{gcd}}^{*} &\ \langle [(1, 6)^{\#4}, 9], \left\{(1, 6)\right\}, \emptyset, 2 \rangle \tag{3 $\times$ \textsc{default}}\\
    \trR{\textsc{gcd}} &\ \langle [9], \left\{(1, 6)\right\}, \emptyset, 2 \rangle \tag{\textsc{drop}}\\
    \trR{\textsc{gcd}} &\ \langle [(2, 9)^{\#1}], \left\{(1, 6), (2, 9)\right\}, \emptyset, 3 \rangle \tag{\textsc{activate}}\\
    \trR{\textsc{gcd}} &\ \langle [(2, 9)^{\#2}], \left\{(1, 6), (2, 9)\right\}, \emptyset, 3 \rangle \tag{\textsc{default}}\\
    \trR{\textsc{gcd}} &\ \langle [3, (2, 9)^{\#2}], \left\{(1, 6)\right\}, \left\{\mathbf{(subtract, 1, 2)}\right\}, 3 \rangle \tag{\textsc{apply} (\textit{subtract})}\\
    \trR{\textsc{gcd}} &\ \langle [(3, 3)^{\#1}, (2, 9)^{\#2}], \left\{(1, 6), (3, 3)\right\}, \left\{...\right\}, 4 \rangle \tag{\textsc{activate}}\\
    \trR{\textsc{gcd}}^{*} &\ \langle [(3, 3)^{\#3}, (2, 9)^{\#2}], \left\{(1, 6), (3, 3)\right\}, \left\{...\right\}, 4 \rangle \tag{2 $\times$ \textsc{default}}\\
    \trR{\textsc{gcd}} &\ \langle [3, (3, 3)^{\#3}, (2, 9)^{\#2}], \left\{(3, 3)\right\}, \left\{..., \mathbf{(subtract, 3, 1)}\right\}, 4 \rangle \tag{\textsc{apply} (\textit{subtract})}\\
    \trR{\textsc{gcd}} &\ \langle [(4, 3)^{\#1}, (3, 3)^{\#3}, (2, 9)^{\#2}], \left\{(3, 3), (4, 3)\right\}, \left\{...\right\}, 5 \rangle \tag{\textsc{activate}}\\
    \trR{\textsc{gcd}} &\ \langle [(4, 3)^{\#2}, (3, 3)^{\#3}, (2, 9)^{\#2}], \left\{(3, 3), (4, 3)\right\}, \left\{...\right\}, 5 \rangle \tag{\textsc{default}}\\
    \trR{\textsc{gcd}} &\ \langle [0, (4, 3)^{\#2}, (3, 3)^{\#3}, (2, 9)^{\#2}], \left\{(3, 3)\right\}, \left\{...,\mathbf{(subtract, 3, 4)}\right\}, 5 \rangle \tag{\textsc{apply} (\textit{subtract})}\\
    \trR{\textsc{gcd}} &\ \langle [(6, 0)^{\#1}, (4, 3)^{\#2}, (3, 3)^{\#3}, (2, 9)^{\#2}], \left\{(3, 3), (6, 0)\right\}, \left\{...\right\}, 6 \rangle \tag{\textsc{activate}}\\
    \trR{\textsc{gcd}} &\ \langle [(6, 0)^{\#1}, (4, 3)^{\#2}, (3, 3)^{\#3}, (2, 9)^{\#2}], \left\{(3, 3)\right\}, \left\{...,\mathbf{(zero, 6)}\right\}, 6 \rangle \tag{\textsc{apply} (\textit{zero})}\\
    \trR{\textsc{gcd}}^{*} &\ \langle [(6, 0)^{\#4}, (4, 3)^{\#2}, (3, 3)^{\#3}, (2, 9)^{\#2}], \left\{(3, 3)\right\}, \left\{...\right\}, 6 \rangle \tag{3 $\times$ \textsc{default}}\\
    \trR{\textsc{gcd}} &\ \langle [(4, 3)^{\#2}, (3, 3)^{\#3}, (2, 9)^{\#2}], \left\{(3, 3)\right\}, \left\{...\right\}, 6 \rangle \tag{\textsc{drop}}\\
    \trR{\textsc{gcd}}^{*} &\ \langle [(4, 3)^{\#4}, (3, 3)^{\#3}, (2, 9)^{\#2}], \left\{(3, 3)\right\}, \left\{...\right\}, 6 \rangle \tag{2 $\times$ \textsc{default}}\\
    \trR{\textsc{gcd}} &\ \langle [(3, 3)^{\#3}, (2, 9)^{\#2}], \left\{(3, 3)\right\}, \left\{...\right\}, 6 \rangle \tag{\textsc{drop}}\\
    \trR{\textsc{gcd}} &\ \langle [(3, 3)^{\#4}, (2, 9)^{\#2}], \left\{(3, 3)\right\}, \left\{...\right\}, 6 \rangle \tag{\textsc{default}}\\
    \trR{\textsc{gcd}} &\ \langle [(2, 9)^{\#2}], \left\{(3, 3)\right\}, \left\{...\right\}, 6 \rangle \tag{\textsc{drop}}\\
    \trR{\textsc{gcd}}^{*} &\ \langle [(2, 9)^{\#4}], \left\{(3, 3)\right\}, \left\{...\right\}, 6 \rangle \tag{2 $\times$ \textsc{default}}\\
    \trR{\textsc{gcd}} &\ \langle [], \left\{(3, 3)\right\}, \left\{...\right\}, 6 \rangle \tag{\textsc{drop}}
  \end{align*}}
  \caption{Execution of the Euclidean algorithm \textsc{gcd} implemented in CHR with initial query $[6,9]$}
  \label{fig:chr:gcd:exec}
\end{figure}

\begin{example}[Greatest common divisor executed]\label{ex:chr:gcd:exec}
  \autoref{fig:chr:gcd:exec} demonstrates the refined semantics on the example query $\left[6, 9\right]$ and the Euclidean algorithm program of \autoref{ex:chr:gcd:decorated}.
  Since the program does not contain any propagation rules, we will abbreviate the propagation history and only show it to emphasize which rule was applied to which values.

  First, the value $6$ is activated and introduced to the store.
  Since there are no other values in the store, yet, its pattern index gets incremented until it is dropped.
  Then, the value $9$ is activated, and its pattern index gets incremented once.
  It now matches the pattern $M^{\#2}$ of the rule $subtract$ and with $6$ matched on $N^{\#3}$, the guard evaluates to $\btrue$ as well.
  Hence, the value $9-6 = 3$ is queried and $(2, 9)$ removed from the store.
  This effectively replaces $9$ with $9-6$.
  The value gets activated, its pattern index incremented to $3$ and the rule $subtract$ can be applied again.
  This time, $6-3 = 3$ is queried and $(1, 6)$ removed, replacing $6$ with $6-3$.
  Now again, $3$ is activated with index $4$ and after one \textsc{default} transition the rule $subtract$ fires again, replacing this newly added $3$ with $0$.
  $0$ then gets activated and instantly matches the $0^{\#1}$ pattern.
  Hence, the rule \emph{zero} fires and removes the value $(6,0)$ from the store.
  At this point, all values except $(3,3)$ are no longer alive and no more non-active values are on the query.
  Hence, all values are successively dropped from the query and the execution terminates.
\end{example}
}{}

\begin{figure*}[t]
  {\footnotesize
  \begin{align*}
                       &\ \langle [(a, b), (b, c)], \emptyset, \emptyset, 1 \rangle  \\
    \trR{\left[trans\right]} &\ \langle [(1, (a, b))^{\#1}, (b, c)], \left\{(1, (a, b))\right\}, \emptyset, 2 \rangle \tag{\textsc{activate}} \\
    \trR{\left[trans\right]}^{*} &\ \langle [(1, (a, b))^{\#3}, (b, c)], \left\{(1, (a, b))\right\}, \emptyset, 2 \rangle \tag{2 $\times$ \textsc{default}} \\
    \trR{\left[trans\right]} &\ \langle [(b, c)], \left\{(1, (a, b))\right\}, \emptyset, 2 \rangle \tag{\textsc{drop}} \\
    \trR{\left[trans\right]} &\ \langle [(2, (b, c))^{\#1}], \left\{(1, (a, b)), (2, (b, c))\right\}, \emptyset, 3 \rangle \tag{\textsc{activate}}\\
    \trR{\left[trans\right]} &\ \langle [(a, c), (2, (b, c))^{\#1}], \left\{(1, (a, b)), (2, (b, c))\right\}, \left\{(\mathit{trans}, 1, 2)\right\}, 3 \rangle \tag{\textsc{apply}}\\
    \trR{\left[trans\right]} &\ \langle [(3, (a, c))^{\#1}, (2, (b, c))^{\#1}], \left\{(1, (a, b)), (2, (b, c)), (3, (a, c))\right\}, \left\{(\mathit{trans}, 1, 2)\right\}, 4 \rangle \tag{\textsc{activate}} \\
    \trR{\left[trans\right]}^{*} &\ \langle [(3, (a, c))^{\#3}, (2, (b, c))^{\#1}], \left\{(1, (a, b)), (2, (b, c)), (3, (a, c))\right\}, \left\{(\mathit{trans}, 1, 2)\right\}, 4 \rangle \tag{2 $\times$ \textsc{default}} \\
    \trR{\left[trans\right]} &\ \langle [(2, (b, c))^{\#1}], \left\{(1, (a, b)), (2, (b, c)), (3, (a, c))\right\}, \left\{\mathbf{(trans, 1, 2)}\right\}, 4 \rangle \tag{\textsc{drop}} \\
    \trR{\left[trans\right]}^{*} &\ \langle [(2, (b, c))^{\#3}], \left\{(1, (a, b)), (2, (b, c)), (3, (a, c))\right\}, \left\{(\mathit{trans}, 1, 2)\right\}, 4 \rangle \tag{2 $\times$ \textsc{default}} \\
    \trR{\left[trans\right]} &\ \langle [], \left\{(1, (a, b)), (2, (b, c)), (3, (a, c))\right\}, \left\{(\mathit{trans}, 1, 2)\right\}, 4 \rangle \tag{\textsc{drop}}
  \end{align*}}
  \caption{Demonstration of the effect of the propagation history.}
  \label{fig:chr:trans}
\end{figure*}

\begin{example}[Transitive hull]\label{ex:chr:trans:cont}
  Given the program from \autoref{ex:chr:trans}.
  \autoref{fig:chr:trans} shows the execution of the program with an initial query $[(a, b), (b, c)]$.

  First, $(a, b)$ gets activated and, after a few transitions, dropped, as the rule requires two values to fire.
  Then, $(b, c)$ gets activated and the rule \emph{trans} fires immediately.
  This queries $(a, c)$ and adds the record $(\textit{trans}, 1, 2)$ to the propagation history.
  Since there is no matching partner for $(a, c)$ in the store, the value gets dropped after activation and two \textsc{default} transitions.
  
  Now, $(b, c)$ is active again.
  Without the propagation history, the rule from above could be applied again, as both $(1, (a, b))$ and $(2, (b, c))$ are still in the store.
  However, since the $(\textit{trans}, 1, 2)$ is already recorded in the propagation history, the \textsc{apply} transition can not be applied.
  Hence, the \textsc{default} transition needs to be applied and the value is dropped ultimately.
\end{example}
  \section{FreeCHR}\label{sec:freechr}
FreeCHR \cite{rechenberger2025freechr} was introduced as a framework to formalize and standardize the embedding of CHR into arbitrary programming languages.
The main idea is to model the syntax of programs as a functor within the domain of the host language.
We want to briefly reiterate the foundational definitions including the \emph{very abstract} operational semantics.
For more details, we refer the reader to the original publication.

\begin{definition}[Syntax of FreeCHR programs]\label{def::freechr:syn:set:functor}
    The functor
    \begin{align*}
        \chr_C D =\
            &\tstr \times \flist \tbool^C \times \flist  \tbool^C \times \tbool^{\flist  C} \times (\flist C)^{\flist  C} \sqcup D \times D
    \end{align*}
    describes the syntax of FreeCHR programs.
\end{definition}
The set $\tstr \times \flist \tbool^C \times \flist \tbool^C \times \tbool^{\flist  C} \times (\flist C)^{\flist C}$ is the set of single rules.
The name of a rule is a string in $\tstr$.
The kept and removed head of a rule are sequences of functions in $\flist \tbool^C$ which map elements of $C$ to Booleans, effectively checking individual values for applicability of the rule.
The guard of the rule is a function in $\tbool^{\flist C}$ and maps sequences of elements in $C$ to Booleans, checking all matched values in the context of each other.
Finally, the body of the rule is a function in $(\flist C)^{\flist C}$ and maps the matched values to a list of newly generated values.
The set $D \times D$ represents the composition of FreeCHR programs by an execution strategy, allowing the construction of more complex programs from, ultimately, single rules.
By the structure of $\chr_C$, a $\chr_C$-algebra with carrier $D$ is defined by two functions 
\begin{align*}
\rho :\ \tstr \times \flist \tbool^C \times \flist \tbool^C \times \tbool^{\flist C} \times (\flist C)^{\flist C} \longrightarrow D &&
\nu :\ D \times D \rightarrow D 
\end{align*}
as $(D, \left[\rho, \nu\right])$.
A $\chr_C$-algebra is called an \emph{instance} of FreeCHR.
The free $\chr_C$-algebra
\begin{align*}
    \chrstar = \left(\mu\chr_C, \left[\chrrule, \chrcomp\right]\right)
\end{align*}
with
\begin{align*}
    \mu\chr_C =
        &\;\tstr \times \flist \tbool^C \times \flist \tbool^C \times \tbool^{\flist C} \times (\flist C)^{\flist C}\\
        \sqcup &\;\mu\chr_C \times \mu\chr_C
\end{align*}
and injections \ifthenelse{\boolean{preprint}}{%
\begin{align*}
    rule &: \tstr \times \flist \tbool^C \times \flist \tbool^C \times \tbool^{\flist C} \times (\flist C)^{\flist C} \longrightarrow \mu\chr_C \\
    \chrcomp &: \mu\chr_C \times \mu\chr_C \longrightarrow \mu\chr_C
\end{align*}
}{{$\chrrule$ and $\chrcomp$}}
provides us with an inductively defined representation of programs.
The program from \autoref{ex:chr:gcd:decorated} (without pattern indices) can be expressed in FreeCHR as shown in \autoref{ex:freechr:gcd}.
\begin{example}[Euclidean algorithm (cont.)]\label{ex:freechr:gcd}
    The program $\mathit{gcd} = \mathit{zero} \odot \mathit{subtract}$ with
    \begin{align*}
        \mathit{zero} &= rule(\mathtt{zero}, [], [\lambda n. n = 0], (\lambda n. \btrue), (\lambda n. [])) \\
        \mathit{subtract} &= rule(\mathtt{subtract}, [\lambda n. 0 < n], [\lambda m. 0 < m], (\lambda n\ m. n \leq m), (\lambda n\ m. \left[m-n\right])) 
    \end{align*}
    implements the Euclidean algorithm in FreeCHR.
    $\lambda$-abstractions are used for ad-hoc definitions of functions.
\end{example}

Finally, we want to recall the \emph{very abstract} operational semantics $\omstara$ of FreeCHR.
\begin{definition}[\emph{Very abstract} operational semantics $\omstara$]\label{def::freechr:sem:op:very-abstract}
    The \emph{very abstract} operational semantics of FreeCHR is defined as the labelled transition system
    \begin{align*}
        \omstara = \langle\; \fmultiset C\;,\; \mu\chr_C\;,\; (\freetr{})^{*}\; \rangle
    \end{align*}
    where the transition relation $(\freetr{}) \subset \fmultiset C \times \mu\chr_C \times \fmultiset C$ is defined by the inference rules described in \autoref{fig:freechr:va}.
    The functor $\fmultiset$ maps a set $X$ to the set $\fmultiset X$ of multisets over $X$.

    \begin{figure}[t]
        \begin{align*}
            \begin{prooftree}[small]
                \hypo{S \freetr{p_j} S'}
                \infer1[step]{S \freetr{p_1 \chrcomp ... \chrcomp p_j \chrcomp ... \chrcomp p_k} S'}
            \end{prooftree}
        \end{align*}
        \begin{align*}
            \begin{prooftree}[small]
                \hypo{k_1(c_1) \band ... \band k_{n}(c_{n}) \band r_{1}(c_{n+1}) \band ... \band r_{m}(c_{n+m}) \band g(c_1,...,c_{n+m}) \equiv_{\tbool} \btrue}
                \infer1[apply]{\left\{c_1,  ...,  c_{n+m}\right\} \uplus \Delta S \freetr{\chrrule(N, [k_1, ..., k_{n}], [r_{1}, ..., r_{m}], g, b)} \left\{c_1, ..., c_{n}\right\} \uplus b(c_1,...,c_{n+m}) \uplus \Delta S}
            \end{prooftree}
        \end{align*}
        \caption{Inference Rules for the \emph{very abstract} operational semantics of FreeCHR $\omstara$}
        \label{fig:freechr:va}
    \end{figure}

    \paragraph*{\textbf{Rule selection}}
    The \textsc{step} transition selects a component program $p_j$ from the composite program \linebreak $p_1 \chrcomp ... \chrcomp p_j \chrcomp ... \chrcomp p_k$.
    \paragraph*{\textbf{Rule application}}
    The \textsc{apply} transition applies a rule to the current state of the program if the state contains a unique value for each pattern in the head of the rule and these values satisfy the guard.
\end{definition}

  \section{\emph{Refined} operational semantics for FreeCHR}\label{sec:freechr:refined}
We now introduce the definition of the \emph{refined} operational Semantics $\omstarr$ of FreeCHR.

\begin{definition}[\emph{Refined} operational semantics $\omstarr$]\label{def:freechr:refined}
    The \emph{refined} operational semantics $\omstarr$ for FreeCHR are defined as an LTS 
    \begin{align*}
        \omstarr = \langle\; \Omega_r C\;,\; \mu \chr^{\#}_C\;,\; (\freetrR{})^{+}\;\rangle
    \end{align*}
    where $(\freetrR{}) \subset \Omega_r C \setprod \mu \chr^{\#}_C \setprod \Omega_r C$ is defined by the transitions described in \autoref{fig:freechr:refined}.

    Analogously to $\prgc^{\#}$, $\mu\chr^{\#}_C$ is the set of FreeCHR programs in $\mu\chr_C$ with enumerated patterns (incrementing from right to left and top to bottom).
    The function $\mathit{indices}$ maps programs $p \in \mu\chr^{\#}_C$ to the set of pattern indices occurring in $p$.

    \begin{figure}[t!]
        \begin{center}
        \begin{align*}
            \begin{prooftree}[small]
                \hypo{c \in C}
                \infer1[activate]{%
                    \langle c : Q, S, H, I \rangle
                    \freetrR{p}
                    \langle \left(I, c\right)^{\#1} : Q, \left\{\left(I, c\right)\right\} \uplus S, H, I+1 \rangle
                }
            \end{prooftree}
            &&
            \begin{prooftree}[small]
                \hypo{l \notin \mathit{indices}(p)}
                \infer1[drop]{%
                    \langle \left(i, c\right)^{\#l} : Q, S, H, I \rangle
                    \freetrR{p}
                    \langle Q, S, H, I \rangle
                }
            \end{prooftree}
        \end{align*}

        \begin{align*}
            \begin{prooftree}[small]
                \hypo{%
                    l \in \mathit{indices}(\chrrule(N, k, r, g, b))
                }
                \hypo{%
                    \langle \left(i, c\right)^{\#l}:Q, S, H, I\rangle
                    \freetrR{\chrrule(N, k, r, g, b)}
                    s
                }
                \infer2[select]{%
                    \langle \left(i, c\right)^{\#l}:Q, S, H, I\rangle
                    \freetrR{p_1 \chrcomp ... \chrcomp \chrrule(N, k, r, g, b) \chrcomp ... \chrcomp p_k}
                    s
                }
            \end{prooftree}
        \end{align*}

        \begin{align*}
            \begin{prooftree}[small]
                \hypo{(i_a, c_a) \in K \uplus R}
                \hypo{h_1(c_1) \wedge ... \wedge h_a(c_a) \wedge ... \wedge h_{n+m}(c_{n+m}) \wedge g(c_1, ..., c_{n+m}) \equiv_{\tbool} \btrue}
                \hypo{(N, i_1, ..., i_{n+m}) \notin H}
                \infer3[apply]{%
                    \langle (i_a, c_a)^{\#l_a}:Q, K \uplus R \uplus \Delta S, H, I \rangle
                    \freetrR{\chrrule(N, k, r, g, b)}
                    \langle B \diamond (i_a, c_a)^{\#l_a} : Q, K \uplus \Delta S, H \uplus \left\{(N, i_1, ..., i_{n+m})\right\}, I \rangle
                }
            \end{prooftree}
        \end{align*}
        {\small with
        $B = b(c_1, ...,c_a, ..., c_{n+m})$,
        $K = \left\{(i_1, c_1), ..., (i_{n}, c_{n})\right\}$, 
        $R = \left\{(i_{n+1}, c_{n+1}), ..., (i_{n+m}, c_{n+m})\right\}$, 
        $K \uplus R = \left\{(i_1, c_1), ..., (i_a, c_a), ..., (i_{n+m}, c_{n+m})\right\}$
        and $k \diamond r = [h_1^{\#l_1}, ..., h_a^{\#l_a} , ..., h_{n+m}^{\#l_{n+m}}]$}

        \begin{align*}
                \begin{prooftree}[small]
                    \hypo{\forall
                        (K \uplus R) \subseteq S.
                            ( h_1(c_1) \wedge ... \wedge h_a(c_a) \wedge ... \wedge h_{n+m}(c_{n+m}) \wedge g(c_1, ..., c_{n+m}) \equiv_{\tbool} \bfalse) \vee (N, i_1, ..., i_{n+m}) \in H}
                    \infer1[default]{%
                        \langle (i_a, c_a)^{\#l_a}:Q, S, H, I \rangle
                        \freetrR{\chrrule(N, k, r, g, b)}
                        \langle (i_a, c_a)^{\#l_a+1} : Q, S, H, I \rangle
                    }
                \end{prooftree}
        \end{align*}
        {\small with $K \uplus R = \left\{(i_1, c_1), ..., (i_a, c_a), ..., (i_{n+m}, c_{n+m})\right\}$ and $k \diamond r = [h_1^{\#l_1}, ..., h_a^{\#l_a} , ..., h_{n+m}^{\#l_{n+m}}]$}
        \end{center}
        \caption{Inference Rules for the \emph{refined} operational semantics of FreeCHR $\omstarr$}
        \label{fig:freechr:refined}
    \end{figure}    

    \paragraph*{\textbf{Activate}}
    The \textsc{activate} transition is a direct translation of the \textsc{activate} rule of $\omega_r$.
    It introduces the value together with a unique identifier to the store and decorates the value on the query with that identifier as well as an initial pattern index.

    \paragraph*{\textbf{Drop}}
    The \textsc{drop} transition is the direct translations of the \textsc{drop} rule of $\omega_r$.
    It removes the currently active value from the query if its pattern index exceeds the indices of the program.

    \paragraph*{\textbf{Select}}
    The \textsc{select} transition initiates the \textsc{apply} or \textsc{default} transitions by selecting the rule \linebreak$rule(N, k, r, g, b)$ from the composition $p_1 \chrcomp ... \chrcomp \chrrule(N, k, r, g, b) \chrcomp ... \chrcomp p_k$ which contains the pattern index of the currently active value.

    \paragraph*{\textbf{Apply}}
    The \textsc{apply} transition, is a translation of the \textsc{apply} transition of $\omega_r$.
    It applies the rule \linebreak$\chrrule(N, k, r, g, b)$ to the state if $c_a$ satisfies the pattern $h_a$ (which is determined by the label index $l_a$),
    the store contains a pair $(i_{\iota}, c_{\iota})$ for every other pattern $h_{\iota}$, such that $h_{\iota}(c_{\iota})$ evaluates to $\btrue$,
    the guard $g(c_1, ..., c_{n+m})$ evaluates to $\btrue$,
    and the configuration $(N, i_1, ..., i_{n+m})$ is not already recorded in the propagation history.
    If all conditions are met, the values $\left\{(i_{n+1}, c_{n+1}), ..., (i_{n+m}, c_{n+m})\right\}$ are removed from the store,
    the values $B$ generated by the body $b(c_1, ..., c_{n+m})$ are queried
    and the configuration $(N, i_1, ..., i_{n+m})$ is recorded in the propagation history, to prevent reapplication.

    \paragraph*{\textbf{Default}}
    Finally, the transition
     is a translation of the \textsc{default} rule of $\omega_r$.
    It can be applied if no other transition is applicable.
\end{definition}
   
The \emph{refined} semantics $\omega_r^{\star}$ for FreeCHR are mostly a direct translation of the \emph{refined} semantics for CHR and operate on the same kind of states.
The only major difference is that the transition rule \textsc{step} provides a proxy for \textsc{apply} and \textsc{default}.
\textsc{step} demands to select a rule which has a pattern with the pattern index of the currently active value.

  \section{Correctness w.r.t. the very abstract operational semantics of FreeCHR}\label{sec:freechr:refined:va}
We now want to establish the \emph{refined} semantics $\omega_r^{\star}$ as a valid concretization of the \emph{very abstract} semantics $\omstara$.
\ifthenelse{\boolean{submission}}{The full proofs of this section and of \autoref{sec:freechr:refined:refined} can be found in the preprint\cite{rechenberger2025refined}.}{}

Since the semantics are defined on different domains of states, we first define a function which strips the additional decoration needed by the \emph{refined} semantics.
\begin{definition}[Abstraction function]
    The function
    \begin{align*}
        &\abstractr : \Omega_r C \longrightarrow \fmultiset C\\
        &\abstractr\langle Q, S, \_, \_ \rangle = \left\{c \mid (\_, c) \in S\right\} \uplus \left\{c \mid c \in Q \cap C\right\}
    \end{align*}
    transforms the states of $\Omega_r C$ into multisets in $\fmultiset C$.
\end{definition}
We use a similar approach to the proof of soundness of $\omr$ \WRT the \emph{theoretical} operational semantics $\omega_t$ \cite{duck2004refined}.
Additionally, the function \ifthenelse{\boolean{preprint}}{%
    \begin{align*}
        &\mathit{unenum} : \mu\chr^{\#}_C \longrightarrow \mu\chr_C\\
        &\mathit{unenum}(\chrrule(N, [h_1^{\#l_1}, ..., h_n^{\#l_n}], [h_{n+1}^{\#l_{n+1}}, ..., h_{n+m}^{\#l_{n+m}}], g, b))
        = \chrrule(N, [h_1, ..., h_n], [h_{n+1}, ..., h_{n+m}], g, b)\\
        &\mathit{unenum}(p_1 \chrcomp p_2) = \mathit{unenum}(p_1) \chrcomp \mathit{unenum}(p_2)
    \end{align*}
    }{$\mathit{unenum}$} removes all pattern indices from a program.

\begin{theorem}\label{th:freechr:soundness:very_abstract}
    The \emph{refined} operational semantics of FreeCHR $\omstarr$ is $(\abstractr, \mathit{unenum})$-sound \WRT the \emph{very abstract} operational semantics $\omstara$.
\end{theorem}

\ifthenelse{\boolean{preprint}}{%
\begin{proof}
    In order to prove \autoref{th:freechr:soundness:very_abstract}, we first show 
    \begin{align*}
        s \freetrR{p} s' \in (\freetrR{}) \Longrightarrow \abstractr(s) \freetr{\mathit{unenum}(p)} \abstractr(s') \in (\freetr{})^{*} 
    \end{align*} via induction over the inference rules that define $(\freetrR{})$.
    \autoref{th:freechr:soundness:very_abstract} then follows from the properties of the reflexive-transitive closure.

    \begin{indbase}[\textsc{activate}]
        Given a proof
        \begin{align*}
            \begin{prooftree}[small]
                \hypo{c \in C}
                \infer1[activate]{%
                    \langle c : Q, S, H, I \rangle
                    \freetrR{p}
                    \langle \left(I, c\right)^{\#1} : Q, \left\{\left(I, c\right)\right\} \uplus S, H, I+1 \rangle
                }
            \end{prooftree}
        \end{align*}
        we know that the reflexive element
        \begin{align*}
            \abstractr\langle c : Q, S, H, I \rangle \freetr{p}
            \abstractr\langle \left(I, c\right)^{\#1} : Q, \left\{\left(I, c\right)\right\} \uplus S, H, I+1 \rangle
        \end{align*}
        is in $(\freetr{})^{*}$ since
        \begin{align*}
            &\abstractr\langle v : Q, S, H, I \rangle\\
            =\ &\lefteqn{\overbrace{\phantom{\left\{c \mid c \in Q, c \in C\right\} \cup \left\{v\right\}}}^{=\left\{c \mid c \in (v:Q), c \in C\right\}}} \left\{c \mid c \in Q, c \in C\right\} \cup\underbrace{\left\{v\right\} \cup \left\{ c \mid (\_, c) \in S \right\}}_{=\left\{ c \mid (\_, c) \in \left\{\left(I, v\right)\right\} \uplus S \right\}}\\
            =\ &\abstractr\langle \left(I, v\right)^{\#1} : Q, \left\{\left(I, v\right)\right\} \uplus S, H, I+1 \rangle
        \end{align*}
        \hfill\checkmark
    \end{indbase}

    \begin{indbase}[\textsc{drop}]
        Given a proof
        \begin{align*}
            \begin{prooftree}[small]
                \hypo{l \notin \mathit{indices}(p)}
                \infer1[drop]{%
                    \langle \left(i, c\right)^{\#l} : Q, S, H, I \rangle
                    \freetrR{p}
                    \langle Q, S, H, I \rangle
                }
            \end{prooftree}
        \end{align*}
        we know that the reflexive element
        \begin{align*}
                \abstractr\langle \left(i, c\right)^{\#l} : Q, S, H, I \rangle \freetr{p}
                \abstractr\langle Q, S, H, I \rangle
        \end{align*}
        is in $(\freetr{})^{*}$ since
        \begin{align*}
            \abstractr\langle \left(i, c\right)^{\#l} : Q, S, H, I \rangle
            = \left\{c \mid c \in Q \band c \in C \right\} \cup \left\{ c \mid \left(\_, c\right) \in S \right\}
            = \abstractr\langle Q, S, H, I \rangle
        \end{align*}
        \hfill\checkmark
    \end{indbase}

    \begin{indbase}[\textsc{apply}]
        Given a proof
        \begin{align*}
            \begin{prooftree}[small]
                \hypo{(i_a, c_a) \in K \uplus R}
                \hypo{P \wedge g(c_1, ..., c_{n+m}) \equiv_{\tbool} \btrue}
                \hypo{(N, i_1, ..., i_{n+m}) \notin H}
                \infer3[apply]{%
                    \langle (i_a, c_a)^{\#l_a}:Q, K \uplus R \uplus \Delta S, H, I \rangle
                    \freetrR{p}
                    \langle B \diamond (i_a, c_a)^{\#l_a} : Q, K \uplus \Delta S, H \uplus \left\{(N, i_1, ..., i_{n+m})\right\}, I \rangle
                }
            \end{prooftree}
        \end{align*}
        with $p = \chrrule(N, k, r, g, b)$ and
        \begin{align*}
            &P = h_1^{\#l_1}(c_1) \wedge ... \wedge h_a^{\#l_a}(c_a) \wedge ... \wedge h_{n+m}^{\#l_{n+m}}(c_{n+m}) &
            &B = b(c_1, ...,c_a, ..., c_{n+m})\\
            &K = \left\{(i_1, c_1), ..., (i_{n}, c_{n})\right\} &
            &R = \left\{(i_{n+1}, c_{n+1}), ..., (i_{n+m}, c_{n+m})\right\} \\
            &K \uplus R = \left\{(i_1, c_1), ..., (i_a, c_a), ..., (i_{n+m}, c_{n+m})\right\} &
            &k \diamond r = [h_1^{\#l_1}, ..., h_a^{\#l_a} , ..., h_{n+m}^{\#l_{n+m}}]
        \end{align*}
        Let 
        \begin{align*}
            &K' = \left\{c_1, ..., c_{n}\right\} &
            &R' = \left\{c_{n+1}, ..., c_{n+m}\right\}\\
            &B' = \left\{c\ |\ c \in B \right\} &
            &\Delta S' = \left\{q\ |\ q \in Q, c \in C\right\} \cup \left\{c\ |\ (\_, c) \in \Delta S\right\}
        \end{align*}
        From $P \equiv_{\tbool} \btrue$ and $g(c_1, ..., c_{n+m}) \equiv_{\tbool} \btrue$, we know that
        \begin{align*}
            h_1(c_1) \wedge ... \wedge h_a(c_a) \wedge ... \wedge h_{n+m}(c_{n+m}) \wedge g(c_1, ..., c_{n+m}) \equiv_{\tbool} \btrue
        \end{align*}
        is true as well.
        Furthermore, from the definition of $\abstractr$, we also know that 
       \begin{align*}
            &\abstractr\langle (i_a, c_a)^{\#l_a}:Q, K \uplus R \uplus \Delta S, H, I \rangle\\
            =\ &\underbrace{\left\{k \mid k \in K \right\}}_{= K'} \cup \underbrace{\left\{r \mid r \in R \right\}}_{=R'} \cup \underbrace{\left\{c \mid c \in Q, c \in C\right\} \cup \left\{s \mid s \in \Delta S \right\}}_{=\Delta S'} \\
            =\ &K' \cup R' \cup \Delta S'
        \end{align*}
        and
        \begin{align*}
            &\abstractr\langle B \diamond (i_a, c_a)^{\#l_a} : Q, K \uplus \Delta S, H \uplus \left\{(N, i_1, ..., i_{n+m})\right\}, I \rangle\\
            =\ &\left\{k \mid k \in K \right\} \cup \left\{c \mid c \in B \diamond Q, c \in C\right\} \cup \left\{s \mid s \in \Delta S \right\}\\
            =\ &\left\{k \mid k \in K \right\} \cup \underbrace{\left\{c \mid c \in B\right\}}_{= B'} \cup \left\{c \mid c \in Q, c \in C\right\} \cup \left\{s \mid s \in \Delta S \right\}\\
            =\ &K' \cup B' \cup \Delta S'
        \end{align*}
        We can hence construct a proof
        \begin{align*}
            \begin{prooftree}[small]
                \hypo{h_1(c_1) \wedge ... \wedge h_a(c_a) \wedge ... \wedge h_{n+m}(c_{n+m}) \wedge g(c_1, ..., c_{n+m}) \equiv_{\tbool} \btrue}
                \infer1[apply]{
                    K' \uplus R' \uplus \Delta S' \freetr{\mathit{unenum}(p)} K' \uplus B' \uplus \Delta S'
                }
            \end{prooftree}
        \end{align*}
        \hfill\checkmark
    \end{indbase}

    \begin{indbase}[\textsc{default}]
        Given a proof
        \begin{align*}
            \begin{prooftree}[small]
                \hypo{\forall
                    (K \uplus R) \subseteq S.
                        ( P \wedge g(c_1, ..., c_{n+m}) \equiv_{\tbool} \bfalse) \vee (N, i_1, ..., i_{n+m}) \in H}
                \infer1[default]{%
                    \langle (i_a, c_a)^{\#l_a}:Q, S, H, I \rangle
                    \freetrR{\chrrule(N, k, r, g, b)}
                    \langle (i_a, c_a)^{\#l_a+1} : Q, S, H, I \rangle
                }
            \end{prooftree}
        \end{align*}
        we know that the reflexive element 
        \begin{align*}
            \abstractr\langle (i_a, c_a)^{\#l_a}:Q, S, H, I \rangle \freetr{\mathit{unenum}(\chrrule(N, k, r, g, b))}
            \abstractr\langle (i_a, c_a)^{\#l_a+1} : Q, S, H, I \rangle
        \end{align*}
        is in $(\freetr{})^{*}$ since
        \begin{align*}
            \abstractr\langle (i_a, c_a)^{\#l_a}:Q, S, H, I \rangle 
            = \abstractr\langle (i_a, c_a)^{\#l_a+1} : Q, S, H, I \rangle
        \end{align*}
        \hfill\checkmark
    \end{indbase}

    \begin{indstep}[\textsc{select}]
        Given a proof
        \begin{align*}
            \begin{prooftree}[small]
                \hypo{l \in \mathit{indices}(p_j)}
                \hypo{%
                    \langle \left(i, c\right)^{\#l}:Q , S, H, I\rangle
                    \freetrR{p_j}
                    s
                }
                \infer2[select]{%
                    \langle \left(i, c\right)^{\#l}:Q , S, H, I\rangle
                    \freetrR{p_1 \chrcomp ... \chrcomp p_j \chrcomp ... \chrcomp p_k}
                    s
                }
            \end{prooftree}
        \end{align*}
        Let $p_1' \chrcomp ... \chrcomp p_j' \chrcomp ... \chrcomp p_k' = \mathit{unenum}(p_1 \chrcomp ... \chrcomp p_j \chrcomp ... \chrcomp p_k)$ and $p_j = rule(...)$.
        With the induction hypothesis
        \begin{align*}
            \langle \left(i, c\right)^{\#l}:Q , S, H, I\rangle \freetrR{p_j} s
            \Longrightarrow
            \abstractr\langle \left(i, c\right)^{\#l}:Q , S, H, I\rangle \freetr{p_j'} \abstractr(s)  \tag{IH}
        \end{align*}
        we can construct a proof
        \begin{align*}
            \begin{prooftree}[small]
                \hypo{\abstractr\langle \left(i, c\right)^{\#l}:Q , S, H, I\rangle \freetr{p_j'} \abstractr(s)}
                \infer1[step]{\abstractr\langle \left(i, c\right)^{\#l}:Q , S, H, I\rangle
                \freetr{p_1' \chrcomp ... \chrcomp p_j' \chrcomp ... \chrcomp p_k'}
                \abstractr(s)}
            \end{prooftree}
        \end{align*}
    \end{indstep}
\end{proof}
}{}
\ifthenelse{\boolean{submission}}{%
\begin{proofsketch}
    We show that 
    \begin{align*}
        \abstractr(s) \freetr{\mathit{unenum}(p)} \abstractr(s') \in (\freetr{})^{*}\ \text{if}\ s \freetrR{p} s' \in (\freetrR{})
    \end{align*}
    via induction over the inference rules of $(\freetrR{})$.
    \autoref{th:freechr:soundness:very_abstract} then follows from the properties of the transitive closure.
    The first base cases \textsc{activate}, \textsc{drop} and \textsc{default} can be reduced to reflexive transitions, since the modifications performed by them are discarded by $\abstractr$.
    The \textsc{apply} case can be reduced to the \textsc{apply} rule of $\omstara$.
    Finally, the induction step \textsc{select} can be reduced to the \textsc{step} transition of $\omstara$.
    
    \hfill\qed
\end{proofsketch}
}{}
With \autoref{th:freechr:soundness:very_abstract}, we established the \emph{refined} semantics $\omstarr$ as a valid concretization of $\omstara$.
Since $\omstara$ is already proven to be sound and complete \WRT $\oma$, \autoref{th:freechr:soundness:very_abstract} already establishes possible implementations of $\omstarr$ as correct implementations of CHR.
To show that such implementations also behave like existing implementations of CHR, we will continue to prove soundness and completeness of $\omstarr$ \WRT the \emph{refined} semantics $\omr$ of CHR.

  \section{Correctness w.r.t. the refined operational semantics of CHR}\label{sec:freechr:refined:refined}
In order to prove correctness of $\omega_r^{\star}$ \WRT $\omega_r$, we first need to embed FreeCHR programs in $\chr_C^{\#}$ into classical CHR programs in $\prgc^{\#}$.
\begin{definition}[Embedding FreeCHR into CHR]
    The embedding function $\fembed^{\#}$ embeds enumerated FreeCHR programs into enumerated CHR programs in $\prgc^{\#}$.
    It is defined as
    \begin{align*}
        \fembed^{\#}&: \mu\chr^{\#}_C \longrightarrow \prgc^{\#}\\
        \fembed^{\#}&(\chrrule(N, [h_1^{\#l_1}, ..., h_n^{\#l_n}], [h_{n+1}^{\#l_{n+1}}, ..., h_{n+m}^{\#l_{n+m}}], g, b))\\
        &\quad = [N @ c_1^{\#l_1}, ..., c_n^{\#l_n} \setminus c_{n+1}^{\#l_{n+1}}, ..., c_{n+m}^{\#l_{n+m}}\\
        &\quad\quad\Leftrightarrow \left. h_1(c_1) \wedge ... \wedge h_{n+m}(c_{n+m}) \wedge g(c_1, ..., c_{n+m}) \mid b(c_1, ..., c_{n+m}) \right.]\\
        \fembed^{\#}&(p_1 \chrcomp ... \chrcomp p_l) = \fembed^{\#}(p_1) \diamond ... \diamond \fembed^{\#}(p_l)
    \end{align*}
\end{definition}
The embedding works by checking the pattern predicates and the guard of the FreeCHR rule in the guard of the CHR rule.

\begin{example}[Embedding FreeCHR into CHR]
    If we apply the embedding $\fembed^{\#}$ to the enumerated version of the program $gcd$ (including the evaluation of function applications), we get the program
    \begin{align*}
      zero\ &@\ n^{\#1}\ \Leftrightarrow\ n = 0 \band \btrue\ |\ \emptyset \\
      subtract\ &@\ n^{\#3}\ \setminus\ m^{\#2}\ \Leftrightarrow\ 0 < n \band 0 < m \band n \leq m\ |\ m - n
    \end{align*}
    with universally quantified variables $n$ and $m$.
    The program corresponds to the \emph{head-normalization} \cite{duck2005compilation} of the program in \autoref{ex:freechr:gcd}.
\end{example}

\begin{theorem}\label{th:freechr:soundness:refined}
    The \emph{refined} operational semantics $\omstarr$ of FreeCHR is $(\id_{\Omega_r C}, \fembed^{\#})$-sound \WRT the \emph{refined} operational semantics $\omr$ of CHR.
\end{theorem}
\ifthenelse{\boolean{preprint}}{%
\begin{proof}
    We first show
    \begin{align*}
        S \freetrR{p} S' \in (\freetrR{}) \Longrightarrow S \trR{\fembed^{\#}(p)} S' \in (\trR{})^{+}
    \end{align*}
    by induction over the inference rules defining $(\freetrR{})$. \autoref{th:freechr:soundness:refined} then follows from the properties of the transitive closure.

    \begin{indbase}[\textsc{activate}]
        Given a proof
        \begin{align*}
            \begin{prooftree}[small]
                \hypo{c \in C}
                \infer1[activate]{%
                    \langle c : Q, S, H, I \rangle
                    \freetrR{p}
                    \langle \left(I, c\right)^{\#1} : Q, \left\{\left(I, c\right)\right\} \uplus S, H, I+1 \rangle
                }
            \end{prooftree}
        \end{align*}
        we can directly translate it to a transition
        \begin{align*}
            \langle c : Q, S, H, I \rangle
            \trR{\fembed^{\#}(p)}
            \langle \left(I, c\right)^{\#1} : Q, \left\{\left(I, c\right)\right\} \uplus S, H, I+1 \rangle 
        \end{align*}\hfill\checkmark
    \end{indbase}

    \begin{indbase}[\textsc{drop}]
        Given a proof
        \begin{align*}
            \begin{prooftree}[small]
                \hypo{l \notin \mathit{indices}(p)}
                \infer1[drop]{%
                    \langle \left(i, c\right)^{\#l} : Q, S, H, I \rangle
                    \freetrR{p}
                    \langle Q, S, H, I \rangle
                }
            \end{prooftree}
        \end{align*}
        we can assume that the label index $l$ also exceeded the indices of $\fembed^{\#}(p)$.
        We can hence translate it to a transition
        \begin{align*}
            \langle \left(i, c\right)^{\#l} : Q, S, H, I \rangle
            \trR{\fembed^{\#}(p)}
            \langle Q, S, H, I \rangle
        \end{align*}
        \hfill\checkmark
    \end{indbase}

    \begin{indbase}[\textsc{apply}]
        Given a proof
        \begin{align*}
            \begin{prooftree}[small]
                \hypo{l \in \mathit{indices}(\chrrule(N, k, r, g, b))}
                \hypo{P \wedge g(c_1, ..., c_{n+m}) \equiv_{\tbool} \btrue}
                \hypo{(N, i_1, ..., i_{n+m}) \notin H}
                \infer3[apply]{%
                    \langle (i_a, c_a)^{\#l_a}:Q, K \uplus R \uplus \Delta S, H, I \rangle
                    \freetrR{p}
                    \langle B \diamond (i_a, c_a)^{\#l_a} : Q, K \uplus \Delta S, H \uplus \left\{(N, i_1, ..., i_{n+m})\right\}, I \rangle
                }
            \end{prooftree}
        \end{align*}
        with $p = \chrrule(N, k, r, g b)$ and
        \begin{align*}
            &P = h_1^{\#l_1}(c_1) \wedge ... \wedge h_a^{\#l_a}(c_a) \wedge ... \wedge h_{n+m}^{\#l_{n+m}}(c_{n+m}) &
            &B = b(c_1, ...,c_a, ..., c_{n+m})\\
            &K = \left\{(i_1, c_1), ..., (i_{n}, c_{n})\right\} &
            &R = \left\{(i_{n+1}, c_{n+1}), ..., (i_{n+m}, c_{n+m})\right\} \\
            &K \uplus R = \left\{(i_1, c_1), ..., (i_a, c_a), ..., (i_{n+m}, c_{n+m})\right\} &
            &k \diamond r = [h_1^{\#l_1}, ..., h_a^{\#l_a} , ..., h_{n+m}^{\#l_{n+m}}]
        \end{align*}
        Since we assume that $P \wedge g(c_1, ..., c_n) \equiv_{\tbool} \btrue$,
        we know for
        \begin{align*}
            \fembed^{\#}(p) =
            \left(N\ @\ v_1^{\#l_1}, ..., v_n^{\#l_n} \setminus v_{n+1}^{\#l_{n+1}}, ..., v_{n+m}^{\#l_{n}} \ \Leftrightarrow\ G \wedge g(v_1, ..., v_{n+m})\ |\ b(v_1, ..., v_{n+m})\right)
        \end{align*}
        with $G = h_1(v_1) \wedge ... \wedge h_{a}(v_{a}) \wedge ... \wedge h_{n+m}(v_{n+m})$that $G \wedge g(v_1, ..., v_{n+m}) \equiv \btrue$.
        We can hence translate it to a transition
        \begin{align*}
            &\langle (i_a, c_a)^{\#l_a}:Q, K \uplus R \uplus \Delta S, H, I \rangle\\
            &\quad\trR{\fembed^{\#}(p)}
            \langle B \diamond (i_a, c_a)^{\#l_a} : Q, K \uplus \Delta S, H \uplus \left\{(N, i_1, ..., i_{n+m})\right\}, I \rangle
        \end{align*}
        \hfill\checkmark
    \end{indbase}

    \begin{indbase}[\textsc{default}]
        Given a proof
        \begin{align*}
            \begin{prooftree}[small]
                \hypo{l_a \in \mathit{indices}(\chrrule(N, k, r, g, b))}
                \hypo{\forall
                    (K \uplus R) \subseteq S.
                        ( P \wedge g(c_1, ..., c_{n+m}) \equiv_{\tbool} \bfalse) \vee (N, i_1, ..., i_{n+m}) \in H}
                \infer2[default]{%
                    \langle (i_a, c_a)^{\#l_a}:Q, S, H, I \rangle
                    \freetrR{\chrrule(N, k, r, g, b)}
                    \langle (i_a, c_a)^{\#l_a+1} : Q, S, H, I \rangle
                }
            \end{prooftree}
        \end{align*}
        with $P = h_1^{\#l_1}(c_1) \wedge ... \wedge h_a^{\#l_a}(c_a) \wedge ... \wedge h_{n+m}^{\#l_{n+m}}(c_{n+m})$ and
        \begin{align*}
            K \uplus R &= \left\{(i_1, c_1), ..., (i_a, c_a), ..., (i_{n+m}, c_{n+m})\right\} & k \diamond r &= [h_1^{\#l_1}, ..., h_a^{\#l_a} , ..., h_{n+m}^{\#l_{n+m}}]
        \end{align*}
        From the premise
        \begin{align*}
            (K \uplus R) \subseteq S.
                        ( P \wedge g(c_1, ..., c_n) \equiv_{\tbool} \bfalse) \vee (N, i_1, ..., i_{n+m}) \in H
        \end{align*}
        we know that for any $C$-instance of
        \begin{align*}
            &\fembed^{\#}(\chrrule(N, k, r, g, b)) =\\
            &\quad\left(N\ @\ v_1^{\#l_1}, ..., v_n^{\#l_n} \setminus v_{n+1}^{\#l_{n+1}}, ..., v_{n+m}^{\#l_{n}} \ \Leftrightarrow\ G \ |\ b(v_1, ..., v_{n+m})\right)
        \end{align*}
        the guard $G = h_1(c_1) \wedge ... \wedge h_{n+m}(c_{n+m}) \wedge g(c_1, ..., c_{n+m})$ does either evaluate to $\bfalse$ or the record $(N, i_1, ..., i_{n+m})$ is already in the propagation history.
        We can hence translate it to a transition
        \begin{align*}
            \langle (i_a, c_a)^{\#l_a}:Q, S, H, I \rangle
            \trR{\fembed^{\#}(\chrrule(N, k, r, g, b))}
            \langle (i_a, c_a)^{\#l_a+1} : Q, S, H, I \rangle
        \end{align*}
        \hfill\checkmark
    \end{indbase}

    \begin{indstep}[\textsc{select}]
        Given a proof
        \begin{align*}
            \begin{prooftree}[small]
                \hypo{l \in \mathit{indices}(p_j)}
                \hypo{%
                    \langle \left(i, c\right)^{\#l}:Q , S, H, I\rangle
                    \freetrR{p_j}
                    S
                }
                \infer2[select]{%
                    \langle \left(i, c\right)^{\#l}:Q , S, H, I\rangle
                    \freetrR{p_1 \chrcomp ... \chrcomp p_j \chrcomp ... \chrcomp p_k}
                    S
                }
            \end{prooftree}
        \end{align*}
        with the induction hypothesis
        \begin{align*}
            \langle \left(i, c\right)^{\#l}:Q , S, H, I\rangle \freetrR{p_j} S \in (\freetrR{})
            \Longrightarrow
            \langle \left(i, c\right)^{\#l}:Q , S, H, I\rangle \trR{\fembed^{\#}(p_j)} S \in (\trR{})^{+}
        \end{align*}
        Since the label index forces us to try to apply the rule $\fembed^{\#}(p_j)$, we can assume from the induction hypothesis that
        \begin{align*}
            \langle \left(i, c\right)^{\#l}:Q , S, H, I\rangle \trR{\fembed^{\#}(p_1 \chrcomp ... \chrcomp p_j \chrcomp ... \chrcomp p_k)} S \in (\trR{})^{+}
        \end{align*}
    \end{indstep}
\end{proof}
}{}
\ifthenelse{\boolean{submission}}{%
\begin{proofsketch}
    We show that 
    \begin{align*}
        S \trR{\fembed^{\#}(p)} S' \in (\trR{})^{+}\ \text{if}\ S \freetrR{p} S' \in (\freetrR{})
    \end{align*}
    by induction over the inference rules defining $(\freetrR{})$.
    \autoref{th:freechr:soundness:refined} then follows from the definition of the transitive closure.
    The first induction base cases \textsc{activate} and \textsc{drop} can be trivially proven by their counterpart.
    For the \textsc{apply} transition, we know that there is a $C$-instance of the embedded rule that meets the conditions of the \textsc{apply} transition of $\omr$ since we know from the premise that the rule fires.
    For the \textsc{default} transition, we show that, if the premises hold, no other transition of the original definition than \textsc{default} can be applied.
    Finally, for the induction step, we derive from the induction hypothesis that we can perform the transition with $\fembed^{\#}(p)$ if we can perform it with $\fembed^{\#}(p_j)$, as we are forced to choose $p_j$ anyway.
    \hfill\qed
\end{proofsketch}
}{}
\autoref{th:freechr:soundness:refined} establishes $\omstarr$ as a valid concretization of $\omr$.
We continue with stating and proving the opposite direction.

\begin{theorem}\label{th:freechr:completeness:refined}
    The \emph{refined} operational semantics $\omstarr$ of FreeCHR is $(\id_{\Omega_r C}, \fembed^{\#})$-complete \WRT the \emph{refined} operational semantics $\omr$ of CHR.
\end{theorem}
\ifthenelse{\boolean{preprint}}{%
\begin{proof}
    We show 
    \begin{align}
        S \freetrR{p} S' \in (\freetrR{})^{+} \Longleftarrow S \trR{\fembed^{\#}(p)} S' \in (\trR{})^{+}
    \end{align}
    by first showing that
    \begin{align*}
        S \freetrR{p} S' \in (\freetrR{})^{+} \Longleftarrow S \trR{\fembed^{\#}(p)} S' \in (\trR{}) 
    \end{align*}
    from which \autoref{th:freechr:completeness:refined} follows from the definition of the transitive closure.

    \begin{case*}[\textsc{activate}]
        Given a transition
        \begin{align*}
            \langle c:Q, S, H, I \rangle \trR{\fembed^{\#}(p)} \langle (I, c)^{\#1} : Q, \left\{\left(I, c\right)\right\} \cup S, H, I+1 \rangle
        \end{align*}
        we know that $c \in C$ and can hence construct a proof
        \begin{align*}
            \begin{prooftree}[small]
                \hypo{c \in C}
                \infer1[activate]{%
                    \langle c : Q, S, H, I \rangle
                    \freetrR{p}
                    \langle \left(I, c\right)^{\#1} : Q, \left\{\left(I, c\right)\right\} \uplus S, H, I+1 \rangle
                }
            \end{prooftree}
        \end{align*}
        \hfill\checkmark
    \end{case*}

    \begin{case*}[\textsc{drop}]
        Given a transition
        \begin{align*}
            \langle (i, c)^{\#j} : Q, S, H, I \rangle \trR{\fembed^{\#}(p)} \langle Q, S, H, I \rangle
        \end{align*}
        we can assume that $\#j$ exceeds the pattern index of the program $\fembed^{\#}(p)$.
        We can hence construct a proof
        \begin{align*}
            \begin{prooftree}[small]
                \hypo{l \notin \mathit{indices}(p)}
                \infer1[drop]{%
                    \langle \left(i, c\right)^{\#l} : Q, S, H, I \rangle
                    \freetrR{p}
                    \langle Q, S, H, I \rangle
                }
            \end{prooftree}
        \end{align*}
        \hfill\checkmark
    \end{case*}

    \begin{case*}[\textsc{apply}]
        Given a transition
        \begin{align*}
            &\langle (i_a, c_a)^{\#l_a} : Q, K \uplus R \uplus S, H, I\rangle\\
            &\quad\trR{\fembed^{\#}(p)}
            \langle B' \diamond ((i_a, c_a)^{\#l_a} : Q), K \uplus S, \left\{(r, i_1, ..., i_{n+m})\right\} \cup H, I \rangle
        \end{align*}
        We can assume to have a rule
        \begin{align*}
            \fembed^{\#}(p_j) = &(N\ @\ v_1^{\#l_1}, ..., v_{n}^{\#l_n}\ \setminus\ v_{n+1}^{\#l_{n+1}}, ..., v_{n+m}^{\#l_{n+m}}\Longleftrightarrow\\
                &\quad\ h_1(v_1) \wedge ... \wedge h_a(v_a) \wedge ... \wedge h_n(v_{n+m}) \wedge g(v_1, ..., v_{n+m})\ |\ b(v_1, ..., v_{n+m})) \in \fembed^{\#}(p)
        \end{align*}
        such that $[v_1^{\#l_1}, ..., v_{n}^{\#l_n}] \diamond [v_{n+1}^{\#l_{n+1}}, ..., v_{n+m}^{\#l_{n+m}}] = [v_1^{\#l_1}, ..., v_{a}^{\#l_a}, ..., v_{n+m}^{\#l_{n+m}}]$.
        We can hence construct a proof
        \begin{align*}
            \begin{prooftree}[small]
                \hypo{l_a \in \mathit{indices}(p_j)}
                \hypo{%
                    S
                    \freetrR{p_j}
                    S'
                }
                \infer2[select]{%
                    S
                    \freetrR{p_1 \chrcomp ... \chrcomp p_j \chrcomp ... \chrcomp p_l}
                    S'
                }
            \end{prooftree}
        \end{align*}
        with $p_j = \chrrule(N, k, r, g, b)$.

        Let
        \begin{align*}
            &P = h_1^{\#l_1}(c_1) \wedge ... \wedge h_a^{\#l_a}(c_a) \wedge ... \wedge h_{n+m}^{\#l_{n+m}}(c_{n+m}) & &B = b(c_1, ...,c_a, ..., c_{n+m})\\
            &K = \left\{(i_1, c_1), ..., (i_{n}, c_{n})\right\} & &R = \left\{(i_{n+1}, c_{n+1}), ..., (i_{n+m}, c_{n+m})\right\} \\
            &K \uplus R = \left\{(i_1, c_1), ..., (i_a, c_a), ..., (i_{n+m}, c_{n+m})\right\} & &k \diamond r = [h_1^{\#l_1}, ..., h_a^{\#l_a} , ..., h_{n+m}^{\#l_{n+m}}]
        \end{align*}
        Since there must be a $C$-instance
        \begin{align*}
            (N\ &@\ c_1^{\#l_1}, ..., c_{n}^{\#l_n}\ \setminus\ c_{n+1}^{\#l_{n+1}}, ..., c_{n+m}^{l_{n+m}}\ \Longleftrightarrow\ G\ |\ B) \in \Gamma_C(\fembed^{\#}(p_j)) 
        \end{align*}
        with
        \begin{align*}
            G =  h_1(c_1) \wedge ... \wedge h_a(c_a) \wedge ... \wedge h_{n+m}(c_{n+m}) \wedge g(v_1, ..., v_{n+m}) \equiv \btrue
        \end{align*}
        we can assume that
        \begin{align*}
            h_1^{\#l_1}(c_1) \wedge ... \wedge h_a^{\#l_a}(c_a) \wedge ... \wedge h_{n+m}^{\#l_{n+m}}(c_{n+m}) \wedge g(v_1, ..., v_{n+m}) \equiv \btrue
        \end{align*}
        Finally, we can assume that \nolinebreak{$\left\{(r, i_1, ..., i_{n+m})\right\} \notin H$} and can hence construct a proof
        \begin{align*}
            \begin{prooftree}[small]
                \hypo{P \wedge g(c_1, ..., c_{n+m}) \equiv_{\tbool} \btrue}
                \hypo{(N, i_1, ..., i_{n+m}) \notin H}
                \infer2[apply]{%
                    \langle (i_a, c_a)^{\#l_a}:Q, K \uplus R \uplus \Delta S, H, I \rangle
                    \freetrR{p_j}
                    \langle B \diamond (i_a, c_a)^{\#l_a} : Q, K \uplus \Delta S, H \uplus \left\{(N, i_1, ..., i_{n+m})\right\}, I \rangle
                }
            \end{prooftree}
        \end{align*}
        \hfill\checkmark
    \end{case*}

    \begin{case*}[\textsc{default}]
        Given a transition
        \begin{align*}
            \langle (i_a, c_a)^{\#l_a} : Q, S, H, I \rangle \trR{\fembed^{\#}(p)} \langle (i_a, c_a)^{\#l_a+1} : Q, S, H, I \rangle 
        \end{align*}
        We can assume to have a rule
        \begin{align*}
            \fembed^{\#}(p_j) = &(N\ @\ v_1^{\#l_1}, ..., v_{n}^{\#l_n}\ \setminus\ v_{n+1}^{\#l_{n+1}}, ..., v_{n+m}^{\#l_{n+m}}\Longleftrightarrow\\
                &\quad\ h_1(v_1) \wedge ... \wedge h_a(v_a) \wedge ... \wedge h_n(v_{n+m}) \wedge g(v_1, ..., v_{n+m})\ |\ b(v_1, ..., v_{n+m})) \in \fembed^{\#}(p)
        \end{align*}
        such that $[v_1^{\#l_1}, ..., v_{n}^{\#l_n}] \diamond [v_{n+1}^{\#l_{n+1}}, ..., v_{n+m}^{\#l_{n+m}}] = [v_1^{\#l_1}, ..., v_{a}^{\#l_a}, ..., v_{n+m}^{\#l_{n+m}}]$.
        We can hence construct a proof
        \begin{align*}
            \begin{prooftree}[small]
                \hypo{l_a \in \mathit{indices}(p_j)}
                \hypo{%
                    S
                    \freetrR{p_j}
                    S'
                }
                \infer2[select]{%
                    S
                    \freetrR{p_1 \chrcomp ... \chrcomp p_j \chrcomp ... \chrcomp p_l}
                    S'
                }
            \end{prooftree}
        \end{align*}
        with $p_j = \chrrule(N, k, r, g, b)$.
        Since we assume a \textsc{default} transition, there is either no $C$-instance
        \begin{align*}
            (N\ &@\ c_1^{\#l_1}, ..., c_{n}^{\#l_n}\ \setminus\ c_{n+1}^{\#l_{n+1}}, ..., c_{n+m}^{l_{n+m}}\ \Longleftrightarrow\ G\ |\ B) \in \Gamma_C(\fembed^{\#}(p_j))
        \end{align*}
        with
        \begin{align*}
            G =  h_1(c_1) \wedge ... \wedge h_a(c_a) \wedge ... \wedge h_{n+m}(c_{n+m}) \wedge g(v_1, ..., v_{n+m}) \equiv \btrue
        \end{align*}
        or if there is one, $(N, i_1, ..., i_{n+m}) \in H$.
        We can hence construct a proof
        \begin{align*}
            \begin{prooftree}[small]
                \hypo{\forall
                    (K \uplus R) \subseteq S.
                        ( P \wedge g(c_1, ..., c_{n+m}) \equiv_{\tbool} \bfalse) \vee (N, i_1, ..., i_{n+m}) \in H}
                \infer1[default]{%
                    \langle (i_a, c_a)^{\#l_a}:Q, S, H, I \rangle
                    \freetrR{\chrrule(N, k, r, g, b)}
                    \langle (i_a, c_a)^{\#l_a+1} : Q, S, H, I \rangle
                }
            \end{prooftree}
        \end{align*}
    \end{case*}
\end{proof}
}{}
\ifthenelse{\boolean{submission}}{%
\begin{proofsketch}
    We show that 
    \begin{align*}
        S \freetrR{p} S' \in (\freetrR{})^{+}\ \text{if}\ \trR{\fembed^{\#}(p)} S' \in (\trR{})
    \end{align*}
    from which \autoref{th:freechr:completeness:refined} follows by the properties of the transitive clause.
    The \textsc{activate} and \textsc{drop} transitions are again trivial to verify.
    If we can apply the \textsc{apply} transition for an embedded rule, we know that we can use the \textsc{select} transition to select the original rule.
    Since there is a $C$-instance for the embedded rule, we also know that \textsc{apply} transition of $\omstarr$ is applicable to the original rule. 
    That the conditions are met is derived from the assumed premises.
    Finally, the \textsc{default} transition is verified by assuming the premise that there is no $C$-instance of the embedded rule we currently need to apply.
    With this we use the \textsc{select} transition to select the rule and then the \textsc{default} transition.
    That the condition of the \textsc{default} transition is met follows from the premise.
    \hfill\qed
\end{proofsketch}
}{}

With \autoref{th:freechr:soundness:refined} and \autoref{th:freechr:completeness:refined} we established $\omega_r^{\star}$ and $\omega_r$ as equivalent representations of each other.

  \section{Conclusion}\label{sec:conclusion}
In this paper we presented a definition of the \emph{refined} semantics $\omega_{r}^{\star}$ of FreeCHR.
We proved soundness of our definition \WRT the \emph{very abstract} semantics $\omstara$ of FreeCHR and thereby established $\omega_{r}^{\star}$ as a valid concretization of $\omstara$.
We also proved soundness and completeness of $\omega_{r}^{\star}$ \WRT the original \emph{refined} semantics $\omega_{r}$ of CHR and thereby establishes $\omega_{r}^{\star}$ and $\omega_{r}$ as valid representations of each other and hence a FreeCHR implementation with $\omega_{r}^{\star}$ semantics as a correct implementation of CHR.

This provides the formal foundation for implementations of FreeCHR which are as expressive and behave like the established implementations of CHR.
Defining a host language agnostic \emph{algorithmic} representation of the \emph{refined} semantics is the subject of ongoing work \cite{rechenberger2025instance,rechenberger2025optimized}.
It will serve as a blueprint to implement consistent and correct CHR systems and also a platform for the aggregation of existing and new improvements which can then be applied consistently to existing implementations.
Example implementations of FreeCHR with refined semantics can be found on \emph{GitLab}\footnote{\url{https://gitlab.com/freechr/}}.

Future work will mostly be concerned with improvements by adapting and applying known optimization techniques for CHR (see, \EG \cite{vanweert2010efficient}).
We plan to accompany the optimizations both with benchmarks to validate their effectiveness and proofs of correctness to ensure the formal validity of our approach.
This is especially important, as optimizations are often accompanied by deviation from the formal definition.
This causes a gap between formalism and programming language which FreeCHR is designed to close.

  \section*{Acknowledgements}
  We thank the anonymous reviewers for their constructive and mindful feedback.

  \bibliographystyle{class/eptcs}
  \bibliography{local/bibliography}

\end{document}